\documentclass[usegraphicx,usedcolumn,usenatbib]{mn2e}
\usepackage{graphicx}
\usepackage{txfonts}
\usepackage{rotating}
\usepackage{accents}
\usepackage{float}
\usepackage{appendix}

\newcommand{\apjs}{ApJS}

\newcommand{\Msun}{M$_{\sun}$}

\bibpunct[,]{(}{)}{;}{a}{,}{,}

\begin{document}

\title[Galactic Chemical Evolution]
{Galactic Chemical Evolution: Stellar Yields and the Initial Mass Function}

\author[Moll\'{a} et~al.]
{Mercedes Moll\'{a}$^{1,2}$\thanks{E-mail:mercedes.molla@ciemat.es},
Oscar Cavichia$^{2,3}$, Marta Gavil\'{a}n $^{4}$, and Brad K. Gibson $^{5}$\\
$^{1}$ Departamento de Investigaci\'{o}n B\'{a}sica, CIEMAT,
Avda. Complutense 40. E-28040 Madrid, Spain\\ 
$^{2}$ IAG, Universidade de S\~{a}o Paulo, Rua do Mat\~{a}o, 1226, 05508-090,
S\~ao Paulo-SP, Brazil\\ 
$^{3}$ Departamento de F\'{i}sica, Universidade Federal de Itajub\'{a}, 
Av. BPS, 1303, 37500-903, Itajub\'{a}-MG, Brazil\\ 
$^{4}$ Departamento de F\'{\i}sica Te\'{o}rica, Universidad Aut\'{o}noma de 
Madrid, Cantoblanco, E-28049, Madrid, Spain\\ 
$^{5}$ E.A. Milne Centre for Astrophysics, Dept. of Physics \& Mathematics, 
University of Hull, Hull, HU6~7RX, United Kingdom
}

\date{\today}
\pagerange{\pageref{firstpage}--\pageref{lastpage}} \pubyear{2015}
\maketitle 
\label{firstpage}

\begin{abstract}
We present a set of 144 galactic chemical evolution models applied to a 
Milky Way analogue, computed using four sets of low$+$intermediate star 
nucleosynthetic yields, six massive star yield compilations, and six 
functional forms for the initial mass function. The integrated or true 
yields for each combination are derived. A comparison is made between a 
grid of multiphase chemical evolution models computed with these yield 
combinations and empirical data drawn from the Milky Way's disc, 
including the solar neighbourhood.  By means of a $\chi^2$ methodology,
applied to the results of  these multiphase models, the best combination of stellar yields and initial mass function capable
of reproducing these observations is identified.
\end{abstract}


\section{Introduction}

Chemical evolution models (CEM) were developed initially to understand 
observations such as the local metallicity distribution of G/K-dwarfs 
and the radial gradient of abundances through the disc of late-type 
spirals, including the Milky Way Galaxy (MWG). The basic framework for a 
CEM involves a volume of a galaxy within which gas is assumed to flow, 
both inwards via infall and radial flows, as well as outflows; an 
adopted star formation prescription, coupled with an initial mass 
function (IMF), then allows the calculation of the production rate of 
stars of a given mass, supernovae, and the ejection rate of 
nucleosynthetic products back to the interstellar medium (ISM). The 
latter is often characterised via the use of {\sl stellar yields} and 
the integrated or {\sl true} yields, concepts first introduced by 
\citet{tin80}. While the infall and star formation rates (SFR) are 
essential to create and maintain a certain radial abundance gradient, 
the IMF and the stellar yields define the absolute level observed in a 
region.  It is therefore critical to understand the origin of the 
elements and, in particular, the specific stars from which individual 
chemical elements originate, the quantity returned by each said star, 
and the timescale for their ejection back into the ISM.

Since the seminal work of \cite{b2fh} much has been done to improve our 
understanding of stellar nucleosynthesis. Early work focused on stars 
with metallicities similar to that of the Sun, with contemporary work 
now concerned with spanning the full range of metallicities encountered 
in nature. However, code-to-code differences still result in substantive 
differences in the predicted stellar yields across these mass ranges 
\citep[e.g.][]{GLM97}. Due to the quite different end-lives of massive 
stars, as opposed to low $+$ intermediate-mass stars, stellar evolution 
codes have typically separated their applicability to either those which 
end their lives as Type~II supernovae (SN-II) or those which end their 
lives as white dwarfs.

For massive stars, the total yields of elements are usually provided for 
those originating from the supernova explosions or those originating 
from pre-explosion stellar winds; only rarely are both provided, 
self-consistently. The most frequently used set of massive stellar 
yields (hereinafter {\sl mas}) has been that of \citet[][ hereafter 
WOW]{woo95}; to the elements produced in SN-II (for metallicities 
spanning $Z=0$ to $Z_{\sun}$), WOW added the pre-supernova yields of 
\citet{ww86}, but did not include the contribution from pre-SN-II 
stellar winds.  Later, \citet[][ hereafter PCB]{pcb98} provided massive 
star yields for a range of metallicities, but now taking into account 
both the pre- and post-explosion elemental return rates, including the 
stellar winds and the subsequent effect of this mass loss on the 
evolution of the star and on the ejection of the supernova explosion. 
More recently, \citet{lim03} and \citet[ hereafter both sets referred to 
as CLI]{chi04}, \citet{lim06} and \citet[ hereafter both sets referred 
to as LIM]{lim12}, \citet[ hereafter KOB ]{kob06} and 
\citet{rau02,fro06}, and \citet[ hereafter the joined sets referred to 
as HEG]{heg10} have calculated new massive stars 
yields\footnote{\citet{chi13} also give new stellar yields but only for 
solar metallicity stars and for this reason they are not used here.}. We 
will use all these sets in this work.
 
For the low and intermediate mass star (hereinafter {\sl lim}) yields, 
besides the seminal work of \citet{rv81}, where the effects of 
convective dredge-up and the so-called Hot Bottom Burning processes were 
taken into account, more recent yield compilations have been provided by 
\citet{for97,ven05}, \citet[][ hereafter VHK]{vhk97}, \citet[][ 
hereafter MAR]{mar01}, \citet[][ hereafter GAV]{gav05,gav06}, and 
\citet[][ hereafter KAR]{kar10}. A consequence of an ever-improving 
knowledge of AGB physics, is the reduction in the differences in the 
published yields, in particular for the CNO elements. Thus, the work 
done by \citet{sta07}, centered on the mass loss rates, shows that 
changes in the yield by up to $\sim$80\% can result, but only for 
certain isotopes. On the other hand, \citet{ven09} focus their efforts 
on calculating new yields with significantly improved values of the 
opacity. Finally, \citet{cam08} and \citet{gil13} devote their work to 
the case of extremely metal poor stars, whose final evolutionary 
characteristics are not well known at the present.  Apart from the AGB 
evolution, other authors have emphasised the importance of the nuclear 
reactions and their associated numerical parameters; this is the case 
for KAR and also \citet{cri09}. The former re-derived the yields of 
\citet{kar07} with new values of proton capture.  The main differences 
with previous works reside in the yields of $^{19}$F, $^{23}$Na and 
neutron rich isotopes.\footnote{The problems of $^{19}$F and $^{23}$Na 
over-production for AGB yields are outlined in \citet{renda04} and 
\citet{fenner06}, respectively.} Nevertheless, the CNO yields do not 
change significantly amongst these works.

Other important sets of yields available in the literature, such as 
\citet{siess10}, have not been incorporated into our analysis. These 
authors provide yields for super-AGB stars with masses in the range 
7.5--10.5 M$_{\sun}$ and metallicities between Z=1e-4 and 0.04.  The use 
of these tables implies a change in the mass at which one star explodes 
as a SN-II, $m_{SN-II}$, and, more importantly, it introduces a third 
stellar mass range, instead of the two currently used (for 
low$+$intermediate and high-mass stars).  We prefer for simplicity to 
adopt a constant, metallicity-independent, value of $m_{SN-II}=8 \mbox 
M_{\sun}$, rather than introduce an additional free parameter. We will 
explore the influence of this sort of metallicity-dependent SN-II mass 
limit, coupled with extant super-AGB yields, in the next future.

Concerning the IMF, it is still matter of discussion if it is constant 
for all type of galaxies or if there are differences with environment, 
dependencies upon galactic stellar mass or metallicity, or on the local 
star formation rate.  Many recent works suggest that the IMF depends on 
the SFR and/or metallicity of the regions \citep[e.g.][ and references 
therein]{bek13,con13,dop13,fer13,geh13,las13,mcw13,smi13,wei13}, 
implying in most cases that the IMF might also be variable with time, 
but with disagreement among their results. \citet{cal10} used in a CEM 
an IMF which depends on the embedded cluster mass function, resulting in 
an IMF variable with time, as a function of the SFR. They conclude that 
the best fit to the solar neighbourhood data occurs with an IMF 
resembling the standard one. \citet{and13,pea14} also support an 
invariant IMF for all types of systems.  Regardless of these issues of 
invariance, the classical functional forms for the IMF employed in the 
literature, including those of \citet{sal55,mil79,fer90, kro02,cha03} 
and \citet[][ hereinafter SAL, MIL, FER, KRO, CHA, and MAS, 
respectively]{mas13}, whilst broadly similar, are quantitatively 
different from each other. In this work we will use these six forms, 
under the assumption that they are invariant with time.

There are numerous CEMs in the literature, with important differences in 
their results, even for the case of the MWG for which the observational 
data sets are numerous.  In these works, the selection of the best 
model, and the corresponding free parameters, such as the star formation 
rate efficiency and/or infall rate, is performed, for any galaxy, 
comparing their observational data with a CEM built using a set of 
stellar yields with a given IMF \citep{Gib97,gav05,rom10,car11}.  Then, 
if observations cannot be well-reproduced, it can be claimed that an 
alternate set of yields {\it or} IMF might be necessary \citep{her11}.  
Alternatively, it is possible to compare data with models computed using 
different IMFs to see which of these functions are valid, without 
changing the stellar yields; \citet{rom05} did just that, concluding 
that \citet{kro01}, CHA, and MIL are better at reproducing the empirical 
data, than SAL, or \citet{sca98}; \citet{vin15} analyze
the integrated yields comparing results from different IMFs. However, the 
abundances within a galaxy or region therein, with a given star 
formation history, may be very different if another {\sl combination of 
IMF + stellar yields} were to be used. Both ingredients are equally 
important to define the elemental abundances in a region and the 
corresponding temporal evolution.

In this work, we make use of the multiphase chemical evolution model 
originally applied in \citet{fer92,fer94} and \citet{mol96} to the solar 
region, the Galactic disc, and to other external spiral disks, 
respectively.  In \citet[ hereafter MD05]{md05}, a large grid of models 
for a set of 440 theoretical galaxies was generated. In that work the 
IMF was taken from FER and the stellar yields were from WOW and GAV. In 
addition, the yields from Type~Ia supernovae (SN-Ia) \citep{iwa99} were 
included along with the SN-Ia rate time distribution given by 
\citet{ruiz00}. In \citet{cavichia14}, we also used a similar model to 
that of MD05, applied to the MWG, modified to include bar-driven gas 
inflows, which has the effect of changing the SFR radial profile without 
significantly modifying the elemental abundance pattern.

Our objective in this new work is to compute chemical evolution models 
for the MWG, with the same framework, total mass, molecular cloud and 
star formation efficiencies, and infall prescriptions for all of them, 
but with different combinations of stellar yields for massive stars (6 
sets), low $+$ intermediate mass stars (4 sets), and IMFs (6 functions), 
thus resulting in a final grid of 144 models. Our aim is to identify 
which is the best combination able to reproduce simultaneously the 
greatest number of observational constraints, mainly those pertaining to 
the radial distributions of gas, stars, and elemental abundances, and to 
the evolution of the solar region.

The stellar yields and IMFs employed in our analysis are outlined in 
Section~2. The chemical evolution model is presented in Section 3, along 
with the results of the 144 models.  The selection of the best models is 
in Section~4, making use of a $\chi^{2}$ approach, after comparison with 
the observational data (which are provided in Appendix \S\ref{obs}).  
Section~5 is devoted to our conclusions.

\section{Ingredients: Stellar yields, Initial Mass Function, and Integrated Yields}
\label{yields}
\subsection{Stellar yield sets}

The stellar yield $q_{i}(m)$ is defined as the fraction of the initial 
mass $m$ of a star ejected in the form of freshly synthesised element 
$i$ \citep{pagelbook}
\begin{equation}
 q_{i}(m)=\frac{m_{eje,new,i}}{m}.
\end{equation}
and is related to the total mass of this element $i$, $m_{eje,i}(m)$,
ejected by the star throughout its evolution (including pre-SN stellar
winds) and death, via
\begin{equation} 
m_{eje,i}(m)=m\, q_{i}(m) +(m-m_{rem})\, X_{i,0},
\end{equation} 
where $m_{rem}$ is the mass of the stellar remnant and $X_{i,0}$ is the 
abundance of the element $i$ initially present in the star.

Stellar yields are calculated by the stellar evolution community by 
coupling the evolution of the interior stellar structure with the 
relevant associated nuclear reactions. Such calculations provide the 
mass of each element produced and ejected to the ISM by stars of 
different masses throughout their lifetime. In chemical evolution 
models, the stellar yields are usually divided into two ranges of 
stellar masses: 1) Low and intermediate mass stars, which include those 
stars with masses $m \le 8$\,\Msun; and 2) Massive stars, with $m > 
8$\,\Msun, assuming that this is the minimum mass for stars which end 
their lives as SN-II.

\subsubsection{Low and intermediate mass stellar yields}

The main contribution from low and intermediate mass stars to the 
chemical enrichment is done during the Asymptotic Giant Branch (AGB) 
phase, where the mass-loss, thermal pulses, Third Dredge Up (TDU) 
events, and Hot Bottom Burning (HBB) are taking place. The first 
metallicity-dependent yields used in CEM were those from \citet{rv81}. 
In retrospect, the low mass loss rate adopted by the authors led to the 
need for a very large number of thermal pulses, to ensure reasonable 
remnant masses; the consequence of spending such a long time in the AGB 
phase was that almost of the $^{12}$C was transformed into $^{14}$N.

As our knowledge of stellar evolution improved, newer yields were 
released with more accurate mass loss prescriptions, TDU events, and 
HBB. This is the case for the compilation of VHK, whose yields span a 
wide range of masses and metallicities (see Table~\ref{refs-lim}), 
although still with very significant nitrogen production by stars with 
$m > 4$\,\Msun. Later, armed with new stellar prescriptions, MAR 
calculated stellar yields for stars of masses between 1 and 5\,\Msun. In 
her work, she suggested that stars with masses greater than 5\, \Msun\ 
end their lives as SN-II, thus only stars between 3 and 5\,\Msun\ 
contributed to the nitrogen production. The final result was a small 
excess in $^{12}$C and a paucity of $^{14}$N.\footnote{The impact of AGB 
yield selection, including \citet{rv81}, VHK, and MAR yields, as applied 
to CEM models of the Milky Way halo was explored by \citet{GM97}.}

\citet{gav05} and \citet[][ hereafter GAV]{gav06} published new yields 
for low and intermediate mass stars, with masses up to 8 \Msun\ and a 
range of metallicities (see Table~\ref{refs-lim}). The main point of 
their work was the treatment of $^{12}$C and $^{14}$N, concluding that a 
great amount of $^{12}$C in the ISM was ejected by intermediate stars, 
leading to $^{14}$N yields not as great as VHK, nor as low as MAR, and 
reproducing well the observational constraints related to the time 
evolution of the elemental and relative abundances of C, N and O, 
throughout the disc and halo.\footnote{The evolution N/O and its 
relation to O/H is beyond the scope of this work, but forms the basis of 
studies such as \citet{gav06} and \citet{mol06}.}

\begin{table}
\caption{Characteristics of the low and intermediate mass stellar yields 
used in this work.}
\begin{tabular}{ccccc}
\hline
Set Name & Z & Mass Range & Yield &  Solar \\
         &   &  ($\rm M_{\sun}$) & Format  & Abundances  \\
\hline
 VHK    & 0.001   & 0.8--8  & $q_{i}(m)$ & AG89 \\
        & 0.004   &        &             &      \\
        & 0.008   &        &             &      \\
        & 0.020   &        &             &     \\
 MAR    & 0.004   & 0.8--5 &  $m q_{i}(m)$ & GA91 \\
        & 0.008   &        &           &     \\
        & 0.020   &        &           &     \\
 GAV    & 0.0126  & 0.8--8 & $q_{i}(m)$ and $m_{eje,i}(m)$ & GS98 \\
        & 0.0159  &        &           &     \\
        & 0.0200  &        &           &      \\
        & 0.0250  &        &           &     \\
        & 0.0317  &        &           &      \\
 KAR    & 0.0001  & 1--6   & $q_{i}(m)$ and $m_{eje,i}(m)$ & AG89 \\
        & 0.004   &        &           &     \\
        & 0.008   &        &           &    \\
        & 0.020   &        &           &     \\       
\hline
\end{tabular}
\footnotesize{AG89: \citet{ag89}; GA91: \citet{gre91}; GS98: \citet{gs98}}
\label{refs-lim}
\end{table}

We use the stellar yields from VHK, MAR, GAV, and KAR for low and 
intermediate mass stars. Other excellent, more recent sets, such as 
\citet{cri11} or \citet{lag11} are less useful for our purpose here in 
that they either do not yet provide the full mass spectrum (the former 
compilation) or the CNO elements needed for our current work (the latter 
compilation).

\begin{figure*}
\centering
\includegraphics[width=0.75\textwidth,angle=0]{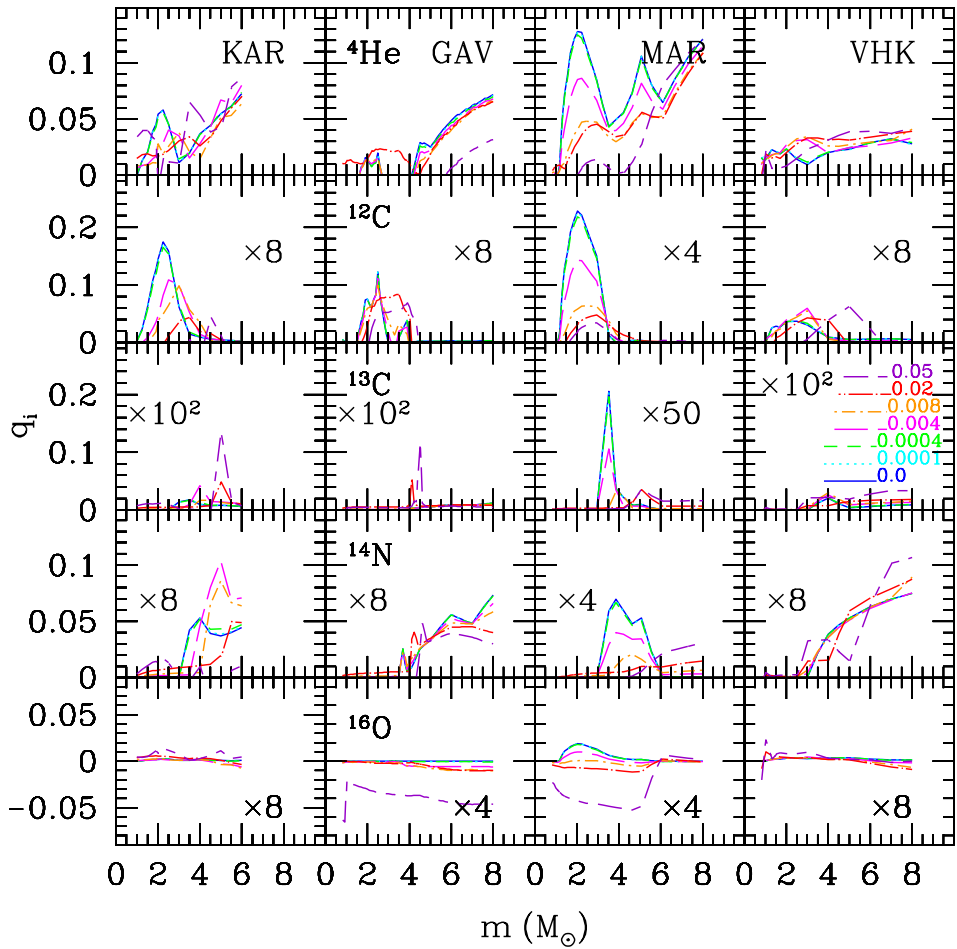}
\caption{Stellar yields for low and intermediate star. Each row shows an 
element, He, $^{12}$C, $^{13}$C, N, and O, from top to bottom, 
and each column refers to a different yield set as labelled.
 The number in each panel is the factor used to multiply the 
yields to plot all of them on a similar y-axis scale. In each panel, 
the coloured lines represent different metallicities as labelled in the 
$\rm ^{13}C$ panel from VHK.  }
\label{qi-lim}
\end{figure*}

From these sets, VHK and MAR give their results as a fraction of stellar 
mass, $q_{i}(m)$, and as mass, $mq_{i}(m)$, respectively, while GAV and 
KAR give both, (net) stellar yields, $m q_{i}(m)$ ,and (total) ejected 
masses $m_{eje,i}(m)$ (see Eq.1). The relationship between both 
quantities depends on the initial abundances $X_{i,0}$, usually assumed 
to be scaled to the solar ones for each value of the total abundance Z
\begin{equation}
 X_{i,0}=X_{i,\sun}\, \frac{Z}{Z_{\sun}},
\end{equation}
while H and He abundances are take to be linear functions with Z:
\begin{eqnarray}
 H& =& H_{p}-\frac{H_{p}-H_{\sun}}{Z_{\sun}}\, Z,\\
 He&=& He_{p}+\frac{He_{\sun}-He_{p}}{Z_{\sun}}\, Z.
\end{eqnarray}

Thus, the initial abundances of each element, $X_{i,0}$ depend on the 
total abundance Z and also on the assumed solar values, $X_{i,\sun}$ and 
for the case of the H and He on the primordial values, $\rm H_{p}$ and 
$\rm He_{p}$, \citep{jim03} as well.  Solar abundances used are 
different for each set, as specified in Table~\ref{refs-lim}; the range 
of masses and metallicities are also listed there.  We have interpolated 
the tables given by these authors to obtain the ejected masses for the 
same 7 metallicities: $Z=0.0$, 0.0001, 0.0004, 0.004, 0.008, 0.02, and 
0.05. We have also normalised the four sets, calculating comparable 
stellar yields $q_{i}(m)$, for each. Table~\ref{yields-lim} gives these 
results for the suite of low and intermediate mass star yields employed 
here.

\begin{table*}
\begin{center}
\caption{Stellar yields $q_{i}(m)$ for our {\sl lim} sets. 
This is an example of the results for VHK and
Z=0.02. The complete Tables 2a to 2d for VHK, MAR, GAV and KAR for the
seven metallicities, are provided in the electronic edition.  We give
for each stellar yield set, the metallicity $Z$, and stellar mass,
$m$, the stellar yields, $q_{i}(m)$, for elements as labelled, 
the remnant mass, $m_{rem}$, and, in the last two columns, the
secondary contributions of $^{13}$C$_{S}$ and $^{14}$N$_{S}$.}
\begin{tabular}{ccccccccccccc}
\hline
$Z$  & $m$ & H & D & $^{3}$He & $^{4}$He & $^{12}$C &$^{13}$C & N & O & $m_{rem}$ &$^{13}\rm C_{S}$ & $\rm ^{14}N_{S}$ \\
   & $M_{\sun}$ & & & & & & & & &  $M_{\sun}$ &  & \\ 
\hline
  0.02 &    3.00  &     -3.86E-02 &      -3.81E-05  &     -2.30E-05 &       3.28E-02 &       5.25E-03 &       6.40E-05 &        1.65E-03  &        4.11E-04   &     0.62 &       4.15E-06  &      1.37E-04 \\        
  0.02 &    3.50  &     -3.84E-02 &      -3.83E-05  &     -2.31E-05 &       3.26E-02 &       5.01E-03 &       6.66E-05 &        1.72E-03  &        3.16E-04   &     0.71 &       3.47E-06  &      1.15E-04 \\        
  0.02 &    4.00  &     -3.75E-02 &      -3.85E-05  &     -2.33E-05 &       3.11E-02 &       4.77E-03 &       6.83E-05 &        1.78E-03  &        1.98E-04   &     0.79 &       2.79E-06  &      9.47E-05 \\        
  0.02 &    4.50  &     -3.78E-02 &      -3.89E-05  &     -2.35E-05 &       3.18E-02 &       2.08E-03 &       1.08E-04 &        4.82E-03  &        1.58E-04   &     0.85 &       2.82E-06  &      9.89E-05 \\        
  0.02 &    5.00  &     -3.85E-02 &      -3.92E-05  &     -2.37E-05 &       3.09E-02 &      -5.94E-04 &       1.40E-04 &        7.24E-03  &       1.42E-04   &      0.92 &       2.71E-06  &      1.04E-04 \\        
\hline
\label{yields-lim}
\end{tabular}
\end{center}
\end{table*}
  
The inferred solar abundances have (in large part) reduced from AG89 to 
the most recent values, such as those from \citet{asp09}.  Since the 
stellar evolution models employed here were constructed with the classic 
solar abundances, the yields must be used assuming that stars have those 
abundances. However, when analysing our results for the solar region, we 
will use the most recent values \citep{asp09}.
  
To compare the different sets, we plot in Fig.~\ref{qi-lim} the stellar 
yields, $q_{i}$, as a function of the stellar mass, for He, $^{12}C$, 
$^{13}C$, N and O.  Although all sets show a broad similarity for each 
element, differences arises when we observe their behaviour in detail. 
As a generic result, MAR differs the most from the others, with a larger 
production for all elements and also a stronger dependence on Z, while 
VHK shows the smallest values.  This is clear in the He panels, as for 
$^{12}$C, for which all sets show a maximum around 4\,$\rm M_{\sun}$ and 
where MAR produces double the quantity of $^{12}$C than KAR or GAV. For 
$^{13}$C, one can see a strong mass-dependence, with an abrupt increase 
for stellar masses only near 3\,$\rm M_{\sun}$.

\begin{figure}
\includegraphics[width=0.35\textwidth,angle=-90]{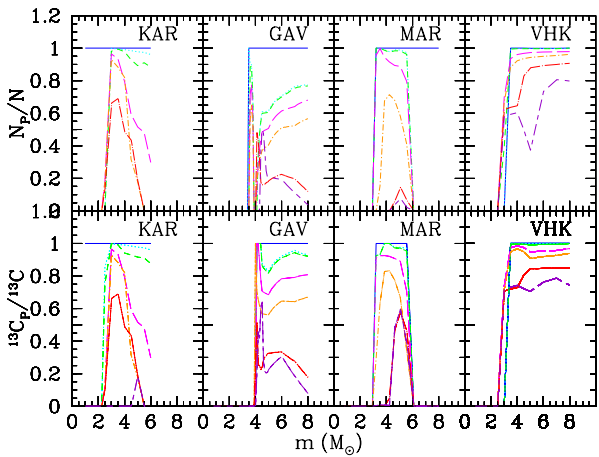}
\caption{Ratio of primary to total stellar yield for $^{13}$C and N: Top 
panels: N$_{\rm P}$/N; Bottom panels: $^{13}$C$_{\rm P}$/$^{13}$C, for {\sl 
lim} sets by KAR, GAV, MAR, and VHK for different metallicities, as 
labelled in Fig.~\ref{qi-lim}.}
\label{PS}
\end{figure}

The stellar yield of N for these low and intermediate mass stars is very 
important since most of N proceeds from this stellar range and because a 
large contribution of the produced N is primary (NP): that is, 
independent of the original metallicity of the star.  This NP is created 
in the HBB process, which needs a minimum core mass to initiate, as it 
occurs with the primary component of the $^{13}$C, as well. N appears 
for stellar masses around 4\,$\rm M_{\sun}$ and when it appears, the 
$\rm^{12}$C consequently decreases. The behaviour is similar for all 
sets, except for MAR, which does not show, unlike the others, the 
increase at the highest mass.  For O, the stellar yield is essentially 
negative (and very low in an absolute sense) for the entire mass range; 
only MAR shows positive values.\footnote{See also \citet{pig13} for 
further recent evidence for modest positive O stellar yields in this 
mass range.} This negative yield will have consequences when the total 
integrated yield for oxygen is used, since the number of stars in this 
mass range is very high, compared with the number of the massive ones 
which produce the bulk of the oxygen.

The contribution of the primary N in galaxies leads to the classical 
relationship between N/O and O/H, in which a clear correlation for high 
metallicity/bright massive galaxies exists, but essentially none for low 
metallicity/low mass systems. Both contributions are separately given, 
or easily obtained, for GAV and MAR stellar yields, but for VHK and KAR 
it was necessary to calculate the primary contribution by the method 
described in \citet{gav06}.  The ratio NP/N is shown in Fig.~\ref{PS} 
for the four sets of low and intermediate stars used in this work.  
Obviously, the ratio NP/N is unity for $Z=0$ and decreases as NS 
(secondary nitrogen) increases with $Z$.

\begin{table}
\caption{Characteristics of the {\sl mas} sets used in this work.}
\begin{tabular}{ccccc}
\hline
Set Name & Z & Mass Range & Mass Loss &  Solar \\
         &   &  (\Msun)   &      &  Abundances \\
\hline
 WOW    & 0.000      & 13--40 & N  &  AG89\\
        & 2\,10$^{-6}$ &      &           &     \\
        & 2\,10$^{-4}$ &      &           &     \\
        & 0.002      &        &           &     \\ 
        & 0.02       &        &           &     \\
 PCB    & 0.0004     & 11-120 &  Y & AG89 \\
        & 0.004      &        &           &     \\  
        & 0.008      &        &           &    \\
        & 0.020      &        &           &     \\
        & 0.050      &        &           &     \\
 CLI    & 0.0000     & 13-35  &  N & AG89 \\
        & 10$^{-6}$  &        &    &     \\
        & 10$^{-4}$  &        &    &     \\
        & 0.001      &        &    &      \\
        & 0.020      &        &    &     \\
 KOB    & 0.000      & 13--40 &  Y &  AG89 \\
        & 0.001      &        &    &                  \\
        & 0.004      &        &    &                 \\
        & 0.020      &        &    &                \\   
 HEG    & 0.000      & 10-100 & Y  &  LO03  \\
        & 0.020      & 12--120&    &    \\
 LIM    & 0.000      & 13-80  & Y  & AG89   \\                  
        & 0.020      & 11-120 &    &         \\  
\hline
\end{tabular}
\footnotesize{AG89: \citet{ag89}; LO03: \citet{lod03}}
\label{refs-mas}
\end{table}

\subsubsection{Massive star stellar yields}

The generation of massive star yields in the literature show the 
deployment of a range of evolutionary codes with different assumptions 
regarding stellar micro-physics, including opacities and nuclear 
reaction rates, and/or macro-physics, such as mixing or mass loss 
prescriptions. The NuGRID collaboration \citep{pig13} has been 
established to rectify this heterogeneous situation, by employing an 
entirely homogeneous micro- and macro-physics approach across the full 
mass and metallicity spectrum (from low- to high-mass stars). However, 
at the time of pursuing this work, the only yields available publicly 
are for masses in the range [1.5--5]\,M$_{\sun}$ and [15-60]\,M$_{\sun}$ 
for Z=0.02 (for Z=0.01, the massive star range reduces to 
[15--25]\,M$_{\sun}$), without including the super-AGB phase and the 
SN-Ia stellar yields. Therefore, while the release of this full grid is 
eagerly anticipated, it is premature to adopt it for these CEMs.

Other well-known stellar yields are those that include a treatment of 
stellar rotation, the internal mixing and structural changes resulting 
from which can lead to appreciable changes in the yields of certain 
elements, in particular nitrogen 
\citep{mey00,mey02a,mey02b,chia06,hir07}. A rich literature now exists 
which examines the role of this stellar rotation on stellar 
nucleosynthesis, although most of them \citep{eks08,mey10,yoon12,chat12} 
have emphasised the impact on very low-metallicity or Population~III 
models.  The lack of an available, fully self-consistent, grid of models 
spanning a range of mass and metallicity (up to solar) has somewhat 
restricted their application for chemical evolution studies. The precise 
treatment of rotational mixing, with velocities varying from 60 to 800 
$\rm km\,^{-1}$ depending upon the authors and codes involved, remains a 
matter of debate.  From a chemical evolution modeling perspective, the 
adoption of a given rotation velocity (and its mass and metallicity 
dependence) implies an additional free parameter, increasing the yield 
options dramatically.\footnote{Besides that, it is not entirely sure 
to what degree, or {\it if}, this rotation is necessary for reproducing 
observations. For example, \citet{tak14} have realized rotating and 
non-rotating models for $Z=0$ for stellar masses between 12 and 140 
M$_{\sun}$. Comparing these models with the three most Fe-deficient 
stars in the Galaxy, they find that abundances for one of them are 
well-reproduced by 50-80 $\mbox M_{\sun}$ non-rotating models, the 
second one is equally well-fitted with non rotating or rotating 15-40 
$\mbox M_{\sun}$ models, and only one of them might require rotating 
30-40 $\mbox M_{\sun}$ models}.  Whilst acknowledging the importance of 
this issue, we feel it premature to proceed with a detailed comparison 
of the rotationally-mixed yields, until a fully self-consistent grid is 
available and calibrated unequivocally with empirical constraints.

\begin{figure*}
\includegraphics[width=0.75\textwidth,angle=0]{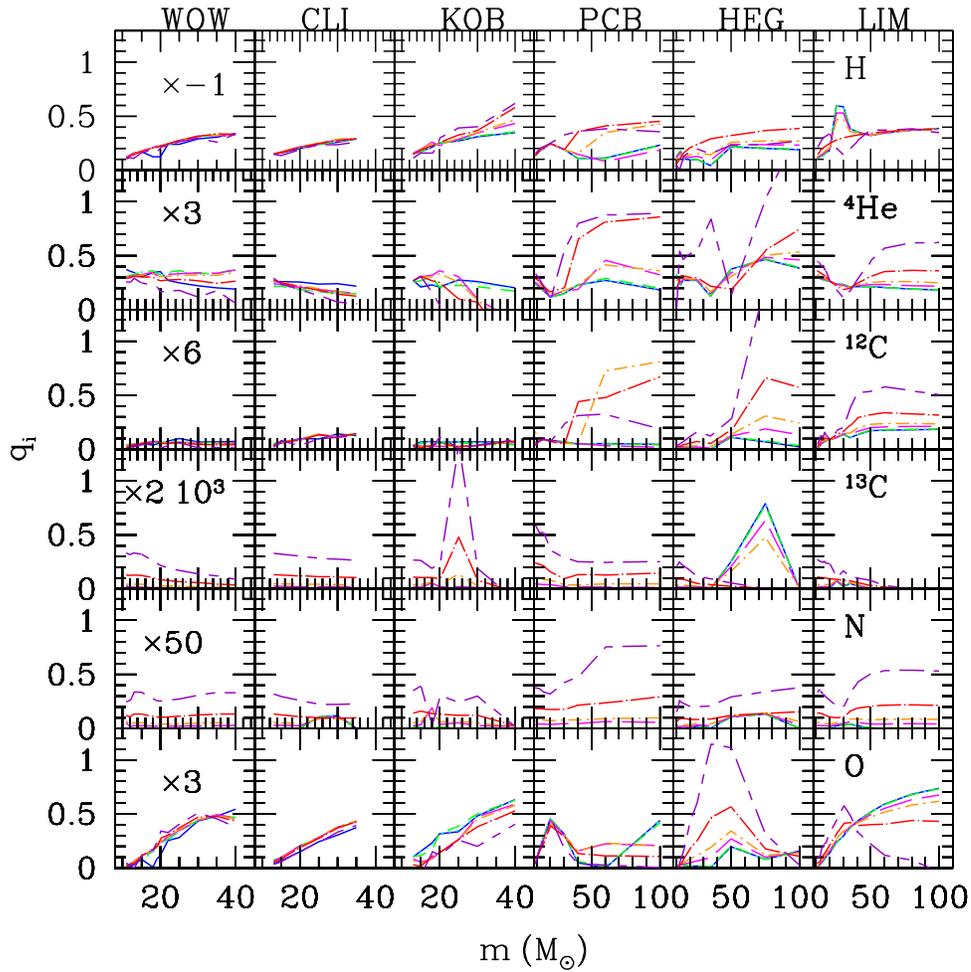}
\caption{Stellar yield for H, He, $^{12}$C, $^{13}$C, N and O for 
massive stars from sets by WOW, CLI, KOB, PCB, HEG, and LIM, for 
different metallicities coded with colours as in Fig.~\ref{qi-lim}.}
\label{qi-mas}
\end{figure*}

Thus, we use the extant compilations in the literature, including the 
sets from WOW, PCB, CLI, HEG, KOB, and LIM. In all cases, authors give 
their results as total ejected masses and most use solar abundances from 
\citet{ag89}, except HEG who use \citet{lod03}. This is given in column 
5 of Table~\ref{refs-mas}, where characteristics of the yields for 
each set are shown: the metallicities, the stellar mass range, and the 
inclusion or not of mass loss.

\begin{table*}
\begin{center}
\caption{Stellar yields $q_{i}(m)$ from different massive stars {\sl mas} sets. 
The values for WOW and Z=0.02 are given here as an example. The 
complete set of tables for WOW, PCB, CLI, KOB, HEG, and LIM, for seven 
metallicities (Z=0, 0.0001, 0.0004, 0.004, 0.008, 0.02, and 0.05) are 
provided in the electronic edition.}
\begin{tabular}{cccccccc}
\hline
$Z$  & $m$ & H & D & $^{3}$He & $^{4}$He & $^{12}$C & $^{16}$O  \\
   & $M_{\sun}$ & & & & & &  \\ 
\hline
  0.02 &   22.00 &    -0.240E+00  &   -0.436E-04  &   -0.701E-06 &     0.908E-01 &     0.823E-02  &       0.100E+00\\   
  0.02 &   25.00 &    -0.270E+00  &   -0.440E-04  &   -0.144E-05 &     0.925E-01 &     0.101E-01  &       0.122E+00\\   
  0.02 &   30.00 &    -0.309E+00  &   -0.449E-04  &   -0.319E-05 &     0.886E-01 &     0.686E-02  &       0.155E+00\\   
  0.02 &   35.00 &    -0.335E+00  &   -0.452E-04  &   -0.472E-05 &     0.801E-01 &     0.632E-02  &       0.158E+00\\   
  0.02 &   40.00 &    -0.331E+00  &   -0.415E-04  &   -0.570E-05 &     0.867E-01 &     0.648E-02  &       0.143E+00\\   
\hline
\label{yields-mas}
\end{tabular}
\begin{tabular}{ccccccccc}
\hline
$^{20}$Ne & $^{24}$Mg & $^{28}$Si & $^{32}$S & $^{40}$Ca & $^{56}$Fe & m$_{\rm rem}$ &$^{13}\rm C_{S}$ & $\rm ^{14}N_{S}$ \\
                    &                     &                    &                 &                    &                     &                          &                                 & \\
\hline
   0.319E-02  &    0.356E-03 &     0.157E-01 &     0.723E-02  &    0.423E-03 &     0.841E-03 &       2.02 &     0.375E-04 &     0.206E-02\\    
   0.158E-01  &   -0.113E-02 &     0.121E-01 &     0.504E-02  &    0.301E-03 &     0.821E-03 &       2.07 &     0.320E-04 &     0.216E-02\\    
   0.144E-01  &    0.782E-02 &     0.100E-01 &     0.272E-02  &    0.692E-04 &     0.787E-03 &       1.94 &     0.265E-04 &     0.244E-02\\    
   0.254E-01  &    0.661E-02 &     0.271E-02 &    -0.132E-03  &   -0.335E-03 &     0.738E-03 &       2.03 &     0.205E-04 &     0.254E-02\\    
   0.310E-01  &    0.429E-02 &     0.916E-03 &    -0.274E-03  &   -0.308E-03 &     0.658E-03 &       5.45 &     0.161E-04 &     0.257E-02\\    
\hline
\end{tabular}
\end{center}
\end{table*}  

\begin{figure}
\includegraphics[width=0.45\textwidth,angle=0]{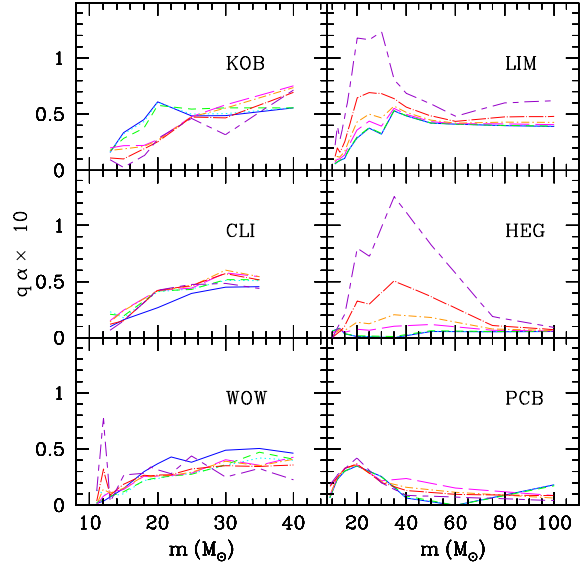}
\caption{Stellar yield for $\alpha$-elements $\rm \alpha=^{20}Ne+ 
^{24}Mg +^{28}Si+^{32}S+^{40}Ca$ for {\sl mas} sets 
for different metallicities. Lines are coded as in the previous figures.}
\label{alfa}
\end{figure}

As noted in Table~\ref{refs-mas}, yields from WOW and CLI do not take 
into account pre-SN stellar winds and their corresponding mass loss, a 
fact particularly important for the most massive stars ($m \ge 30 
M_{\sun}$), and give yields for an upper mass limited to 40 $M_{\sun}$. 
To extrapolate these yields up to 100\,\Msun\ is problematic, since mass 
loss is known to be substantial for these most massive stars. 
Extrapolation without the inclusion of mass loss would result in an 
integrated yield significantly higher than it should be. KOB, PCB, HEG, 
and LIM take into account a treatment of mass loss for these massive 
stars. In the three last sets, yields are provided up to 100-120 \Msun, 
while KOB gives their results up 40 \Msun.  Therefore, we give in 
Table~\ref{yields-mas} the stellar yields up to 40\,\Msun\ for WOW, CLI, 
and KOB, and up to 100\,\Msun\ only for PCB, HEG, and LIM. Graphically, 
we do the same in Figs~\ref{qi-mas} and \ref{alfa}, in three panels (or 
columns of panels) at the left and at the right, respectively

The stellar yields for these sets are given in Table~\ref{yields-mas}. 
We compare, for the same elements as in Fig.~\ref{qi-lim}, the different 
sets of yields of massive stars in Fig.~\ref{qi-mas}.  There, we show 
the results for each element in a row, for each massive star yield set 
in a different column, as labelled at the top of the figure. As a 
generic result, we see a very different behaviour amongst the sets which 
are calculated without taking into account the existence of mass loss by 
stellar winds before the explosion of supernova (e.g., those of WOW and 
CLI), and those which do include these stellar winds (e.g., those of 
PCB, HEG, and LIM).  Curiously, KOB is more similar to the first ones 
although formally this set forms part of the latter.  The first sets are 
in the left-most columns, while the other four are in the right-most 
panels. The ones on the left show a weaker dependence on Z than the ones 
on the right, which is to be expected, as the mass loss is assumed to be 
dependent upon metallicity.  In Fig.~\ref{qi-mas} we plot the yield of 
all elements multiplied by a factor as labelled in the WOW panel, in 
order to compare all of them on a similar scale. In all cases, H is 
negative with values in the range [-0.1,-0.5] depending on the stellar 
mass, on the metallicity, and on the authors.  In LIM there is a strong 
variation around 30\,\Msun, and it becomes positive for the highest 
abundance.  The behaviour for HEG shows quite abrupt changes with mass 
due to the $Z=0$ set.

Since He is produced directly from H, its behavior is complementary to 
that of H, increasing when H decreases, although the absolute value is 
smaller than that, since a certain quantity is necessary to create the 
other elements.  The ejected mass of He and $^{12}$C is higher in the 
case of PCB, HEG, and LIM than those of WOW, CLI, and KOB. When we 
compare the same mass range we see that for $m\le 40$\,\Msun\ more He, 
C, and N is ejected, while O is produced in a smaller quantity when the 
mass loss by stellar winds is included.

For the elements beyond O, the yields are shown in Fig.~\ref{alfa} for 
the six different sets of massive stellar yields. Here, we add the 
yields for elements Ne, Mg, Si, S, and Ca and represent this 
$\alpha$-yield as a function of the stellar mass for each yield set.  As 
for O, a different behaviour arises between the yields calculated taking 
into account the stellar winds (right panels) and those which do not 
(left panels); the former show a maximum around 20-30\,\Msun. KOB shows 
a behaviour between both, similar to those without mass loss, but also 
indicating a slight maximum near 40\,\Msun.

\subsection{The Initial mass function}

\begin{figure}
\includegraphics[width=0.35\textwidth,angle=-90]{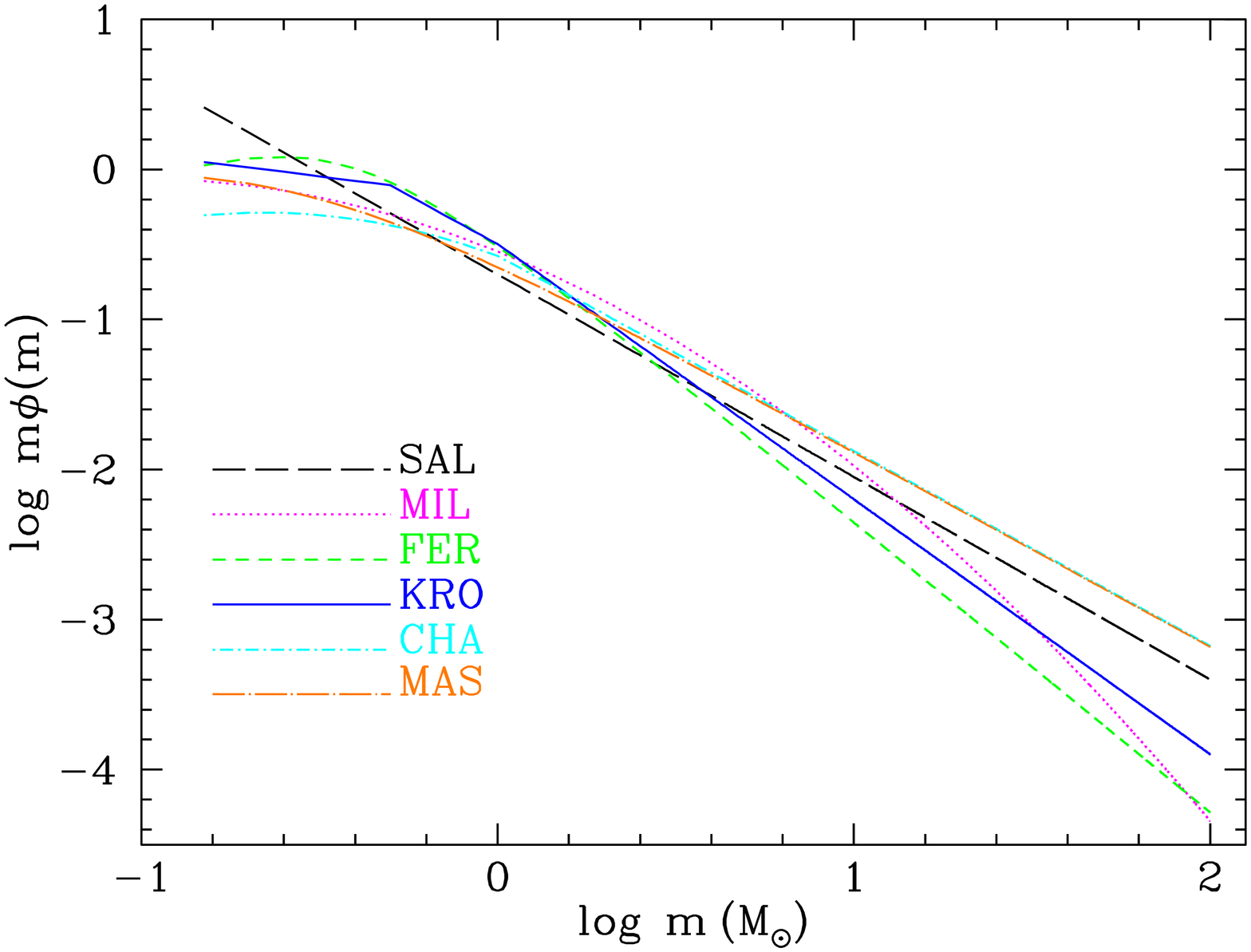}
\caption{The initial mass functions used in this work as $\log{m 
\phi(m)}$ by  SAL, MIL, FER, KRO, CHA, and MAS.}
\label{imf}
\end{figure}

We are building upon the Galactic model outlined in MD05, but instead of 
simply using the FER IMF (as we did in that work), we now employ a range 
of functional forms for the IMF, as well as the various stellar yield 
data sets described in \S2.1. The IMFs adopted are from SAL, MIL, FER, 
KRO, CHA, and MAS, as shown in Fig.~\ref{imf}, where differences amongst 
them appear readily. We assumed the IMF to be invariant with time and 
metallicity.

The functional forms for the adopted IMFs are:
\begin{eqnarray}
\phi(m)_{SAL} & = &  m^{-2.35} \mbox{}, \\
\phi(m)_{MIL} & = & e^{\frac{(\log{m}+1.02)^{2}}{2\,0.68^{2}}},\\
\phi(m)_{FER} & = & 10^{-\sqrt{0.73+\log{m} (1.92+2.07\,\log{m})}}/m^{1.52} \mbox{ }, \\
\phi(m)_{KRO} & = & \left\{ \begin{array}{l r}
  m^{-0.35}    &  \mbox{ $0.15 \le m/M_{\sun} < 0.08$} \\
 0.08 m^{-1.3} &  \mbox{ $0.08 \le m/M_{\sun} < 0.50$} \\
 0.04 m^{-2.3} &  \mbox{ $0.50 \le m/M_{\sun} < 1$ }\\
 0.04 m^{-2.7} &  \mbox{ $ m /M_{\sun} \ge 1 $} \end{array} \right. \\
\phi(m)_{CHA} & = & \left\{ \begin{array}{l r}
0.086 m^{-1} e^{-\frac{(\log{m}+0.657)^{2}}{2\,0.57^{2}}}& \mbox{$0.15 \le m/M_{\sun} < 1$} \\
0.043 m^{-2.3} & \mbox{$1 \le m/M_{\sun} \le 100 $}, \end{array}  \right. \\ 
\phi(m)_{MAS} & = & AA\,\left(\frac{m}{m_{char}}\right)^{-\alpha}\left\{1+\left(\frac{m}{m_{char}}\right)^{1-\alpha}\right\}^{-\beta} \mbox{},\\
\label{imfs}
\end{eqnarray}
where:
$m_{char}=0.2\,\rm M_{\sun}$, 
\begin{equation}
G(m)=\left(1+\left(\frac{m}{m_{char}}\right)^{1-\alpha}\right)^{1-\beta}, \,{\rm and}
\end{equation}
\begin{equation}
AA=\frac{(1-\alpha)\,(1-\beta)}{m_{char}}\,\frac{1}{G(m_{up})-G(m_{low})}
\end{equation}
\begin{table}
\caption{Number of stars for the adopted IMFs for a stellar mass of
10\,$^{4}\rm M_{\sun}$. We show the normalisation constant A, the
total number of stars, $N_{*}$, the number of stars with mass smaller
than 1\,\Msun, $N_{low}$, the number of low and intermediate mass
stars, with $ 4\,\rm{M_{\sun}} \le m \le 8\,\rm{M_{\sun}}$, $N_{lim}$,
the number of massive stars with $m>8$\,\Msun, which will be SN-II,
$N_{SN}$, and the number of stars more massive than 20\,\Msun,
$N_{mas}$. }
\begin{tabular}{ccccccc}
\hline
IMF &    A   & $N_{*}$ & $N_{low}$ &$N_{lim}$ &  $N_{SN}$ & $N_{mas}$ \\
\hline
\multicolumn{7}{c}{$m_{up}=40\,\rm M_{\sun}$}\\
SAL &   2090 & 20200 & 19970 & 145 &   83  &  17 \\
MIL &    191 & 13387 & 13093 & 214 &   79  &   9  \\
FER &  22000 & 18924 & 18792 &  98 &   34  &   5  \\
KRO &  80830 & 17100 & 16925 & 124 &   51  &   8  \\
CHA & 148210 & 11125 & 10780 & 215 &  129  &  26  \\
MAS &  13110 & 13133 & 12800 & 206 &  125  &  25 \\
\hline
\multicolumn{7}{c}{$m_{up}=100\,\rm M_{\sun}$}\\
SAL &   2000 & 19715 & 19490 & 138 &   86  &  23  \\
MIL &    189 & 13329 & 13037 & 212 &   80  &  11   \\
FER &  21869 & 18837 & 18704 &  98 &   35  &   6   \\
KRO &  79458 & 16932 & 16756 & 123 &   54  &  10   \\
CHA & 137808 & 10386 & 10055 & 200 &  131  &  36   \\
MAS &  12222 & 12319 & 11999 & 192 &  128  &  35  \\
\hline
\label{nstar}
\end{tabular}
\end{table}

As usual, the total mass in stars is normalised to 1\,\Msun
\begin{equation}
\int_{m_{low}}^{m_{up}}{A\,m\,\phi(m)\,dm}=1\,M_{\sun}.
\end{equation}
and in this way, the total number of stars, $N_{*}$ in a generation is
\begin{equation}
N_{*}=\int_{m_{low}}^{m_{up}}{A\,\phi(m)\,dm}.
\end{equation}

Our initial plan was to use the same lower ($m_{low}=0.15$~\Msun) and 
upper ($m_{up}=100$~\Msun) mass limits for each CEM; however, as noted 
previously, some yield compilations are restricted to $\le$40~\Msun.  As 
such, we have computed the number of stars for each IMF for these two 
values of $m_{up}$ (see Table~\ref{nstar}) and in the next section, 
models have been computed for each combination of IMF+massive stars with 
a different $m_{up}$ following the set of massive stars used.

\section{Chemical evolution models}
\subsection{Summary Description}

The chemical evolution code used here is that described in MD05 and 
\citet{mol14}, and in \citet[][ hereinafter MCGD]{mcgd}, the latter in 
which we present a new updated grid of chemical evolution models for 
spiral, irregular, and low mass galaxies with some modifications in the 
input parameters over the ones from MD05.

We assume a radial distribution of primordial gas in a spherical 
proto-halo falls onto the plane defining the disc.\footnote{The code is 
inherently one-dimensional (in $r$), involving a thin disc and azimuthal 
symmetry.}  The mass radial distributions are calculated from the 
prescriptions in \citet{sal07}, who give expressions to compute the halo 
density, virial radius, rotation curve, and final mass of the disc as 
functions of the virial mass, $\rm M_{vir}$.  We have calculated an 
initial mass distribution with a dynamical mass of $\sim 10^{12}\,\rm 
M_{\sun}$ and a maximum rotation velocity of $\rm 
V_{rot}=177\,km\,s^{-1}$.  The infall rate or collapse timescale in each 
radial region is chosen in such a way that the disk ends with a radial 
profile similar to that observed, by following the prescriptions from 
\citet{shan06} for the ratio $\rm Mdyn/Mdisk$, at the end of the 
evolution for a time of 13.2\,Gyr. This method gives as a result, for 
the chosen virial mass, a radial distribution of the final mass of the 
disk $M_{D}(R)$ and also the collapse timescale radial distribution, 
$\tau(R)$, necessary to obtain it.

Our formalism for the SFR adopts two stages, first forming molecular 
clouds from the diffuse gas according to a Schmidt law with $n=1.5$, and 
then second, forming stars from cloud-cloud collisions.  Once choosing 
the total mass radial distribution, it is necessary to determine which 
are the best efficiencies to form molecular clouds and stars for this 
MWG-like galaxy. This has been performed in MCGD, comparing the time 
evolution of the region located at R=8\,kpc and the radial distributions 
of gas, stars, and SFR with the present-time data. This comparison 
allowed us to select and fix the best efficiencies to reproduce the MWG 
disk data, which are the one used in this work.

\begin{figure}
\includegraphics[width=0.36\textwidth,angle=-90]{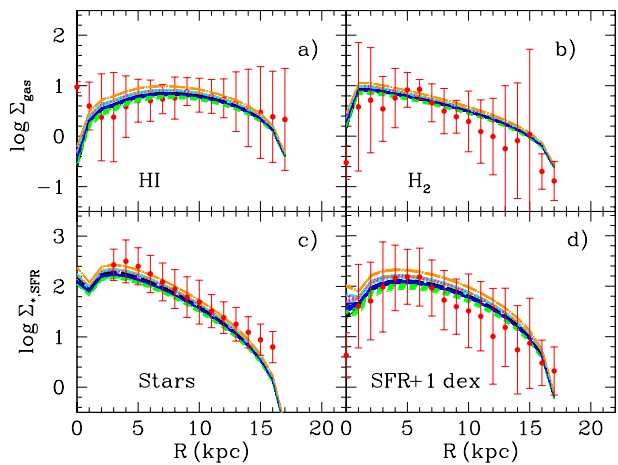}
\caption{The radial distributions of the MWG disk surface densities
for the sets of 144 models compared with the observational data as red
dots with error bars: a) the surface density of diffuse gas,
$\Sigma_{\rm H{\sc i}}$, b) molecular gas, $\Sigma_{H_{2}}$, c) stars,
$\Sigma_{*}$, in units of ${\rm M}_{\sun}\,{\rm pc}^{-2}$ for all of
them, and d) SFR, $\Sigma_{\rmn SFR}$, in units of ${\rm
M}_{\sun}\,{\rm pc}^{-2}\,{\rm Gyr}^{-1}$.  All panels are given in
logarithmic scale. Each colour-type of line indicates a different IMF with
the same coding as in Fig.~\ref{imf}.}
\label{disk}
\end{figure}

For this basic model we have computed all possible combinations of the 
six IMFs with the six {\sl mas} sets and the four {\sl lim} yields 
described in \S\ref{yields}, resulting in a total of 144 models for the 
MWG.  In order to identify the best combination capable of reproducing 
the extant observations, we will now compare the results of these models 
with the observational data given in Appendix \S\ref{obs} where, 
furthermore, the empirical data has been binned in order map these onto 
our model bins.

\subsection{Results for the Solar Vicinity and the MWG disc}

In Fig.~\ref{disk} we present, for the 144 models, the results 
concerning the state of the disc or the radial distributions at the 
present time ($t=13.2$\,Gyr) for gas, stars, and the SFR, compared with 
the data shown in Table~\ref{binned}. For the SFR, panel d), we have 
artificially increased all values (models and data) by 1~dex, in order 
to plot them in the same scale as used in panel c).  The results show a 
small dispersion around the data or around the mean values. These good 
results are the consequence of the infall rate and the star formation 
rate efficiencies selected for the model to reproduce the MWG.  In all 
cases the models' dispersions are comparable to, or smaller than, the 
data uncertainties. These radial distributions are, as expected, only 
slightly dependent upon the IMF, due to the different rate of 
ejected/returned gas by (mostly massive) stars when they die.

\begin{figure}
\includegraphics[width=0.36\textwidth,angle=-90]{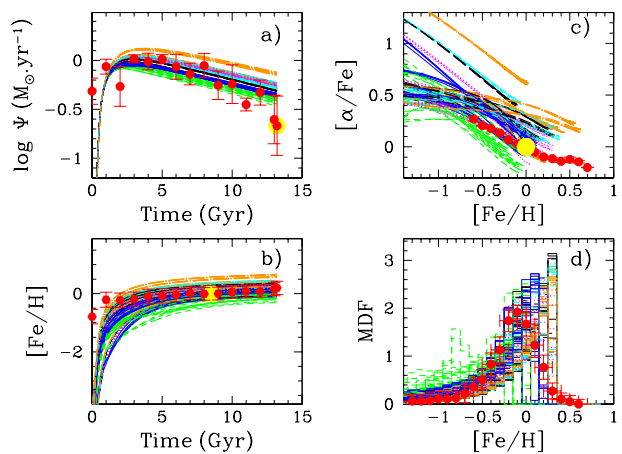}
\caption{The evolution of the Solar Neighbourhood for the set of 144
models compared with observational data as red dots with error bars, as
obtained in \S~\ref{obs_sun}.
The large yellow dot represents the solar values. a) SFR (in
$M_{\sun}\,yr^{-1}$) in logarithmic scale; b) [Fe/H]; c) [$\alpha$/Fe]
{\sl vs} [Fe/H]; d) The MDF.  
The coding of the lines is as in Fig.~\ref{disk}. }
\label{solar}
\end{figure}

The evolution of the SFR, metallicity, [$\alpha$/Fe] as a function of 
[Fe/H], and the Metallicity Distribution Function (MDF) for the solar 
region of our 144 models are compared with the observational data given 
in Tables~\ref{SV-binned} and ~\ref{ofe-binned} in Fig.~\ref{solar}.  It 
is quite evident that, even with the same input parameters and total 
mass for MWG, the resulting evolution is different for each model. In 
the case of the SFR, this is due to the IMF used, since the value of the 
mass locked in stars (1-R) changes with IMF.  Thus, the evolution for 
FER shows the lowest SFR histories while MAS models show the highest, 
since the returned gas fraction of each stellar generation is the lowest 
for the FER models. Within each IMF, each combination {\sl mas+lim} also 
shapes the results somewhat, but in this panel the yields have a smaller 
role than the IMF.  In panel b), the results are the consequence of the 
SFR history, and, therefore, again FER is in the lowest part of the 
model locus while MAS is in the highest.  Since Fe is produced mainly by 
the SN-Ia, the results depend more on IMF than on the stellar yields of 
massive stars; they do not depend upon the low- and intermediate-mass 
stars.

In panel c) we show the classical plot of [$\alpha$/Fe]--[Fe/H] for the 
solar region, where [Fe/H] is often taken as a proxy for time. This 
figure gives the differences in the ejection to the ISM of 
$\alpha$-elements, coming from massive stars, and from the Fe ejected 
mainly by SN-Ia, and also partially due to {\sl mas} yields. Therefore, 
both IMF and massive stellar yields are playing a role here. It is 
evident that a `by eye' inspection of these panels would suggest that 
the KRO, CHA, MIL, and SAL in our models reproduce better the data. When 
we use MAS, results fall above the data for all combinations of stellar 
yields, while our models using FER tend to lie below the observations. 
This plot also gives an indication concerning the massive star yields + 
IMF combinations which may be rejected: WOW is only valid when used with 
FER.  In fact, WOW have already noted that their Fe yield is high and 
recommend it be divided by a factor of two in order to best reproduce 
the data with a CEM. This high Fe yield is compensated for when using 
FER, since the number of massive stars is small in this IMF compared 
with the others. In panel d) it is again evident that the IMF has an 
important effect on the MDF, with most of FER models at the left and MAS 
models at the right of the observations. Again our KRO, MIL, and some 
CHA models seem to fit better the observed MDF.

\begin{figure}
\includegraphics[width=0.45\textwidth,angle=0]{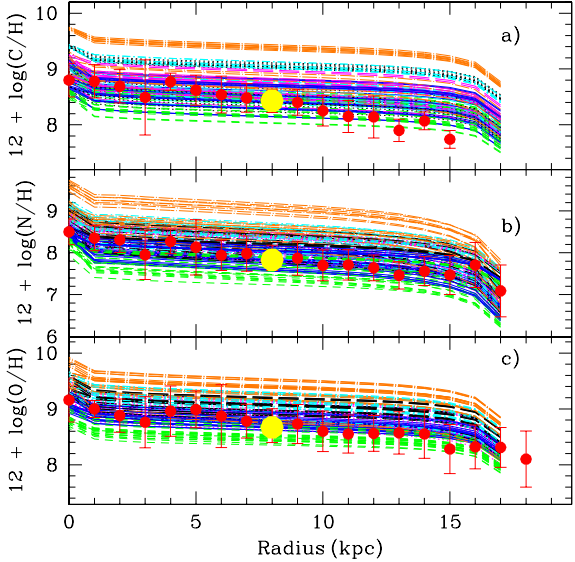}
\caption{The radial distributions of elemental abundances, as $\rm 12
+ log (X/H)$: a) C, b) N and c) O, in the MWG disx for the sets of 144
models, compared with observational data as red dots with error bars.
Coding of the lines is as for Fig.~\ref{solar}.}
\label{abun}
\end{figure}

Finally, we present the resulting radial distributions of C, N, and O for 
the whole set of 144 models in Fig.~\ref{abun}, compared with the binned 
data obtained in \S\ref{abunobs}. The slope of the radial abundance 
gradients does not depend, as expected, on the combination {\sl IMF+ 
stellar yield}. The radial gradient is determined by the ratio between 
SFR and infall rate, $\Psi(t)/f$, and it is basically independent of the 
IMF or the stellar yields.
\begin{table*}
\begin{center}
\caption{Values of $\chi^{2}$ obtained from the fitting of models
to each one of the data sets defined in Appendix~\ref{obs}. 
The entire table is presented in the electronic version. A
portion is shown here for guidance regarding its form and content.}
\label{tabla_chi}
\begin{tabular}{cccrrrrrrrrrrrr}
\hline
{\sl lim} & {\sl mas} & IMF & $\chi^{2}_{\Psi_t}$ & $\chi^{2}_{[Fe/H]}$ & $\chi^{2}_{[\alpha/Fe]}$ & $\chi^{2}_{MDF}$ & $\chi^{2}_{HI}$ &$\chi^{2}_{H_{2}}$ &$\chi^{2}_{*}$ &$\chi^{2}_{\Psi_R}$ &$\chi^{2}_{C/H}$ &$\chi^{2}_{N/H}$ &$\chi^{2}_{O/H}$ \\
\hline
GAV& CLI & SAL &   11.768 &   2.387  & 	 70.848  &   92.836 & 1.623 &   5.157 &   2.517 &   2.805 &   51.846 &    7.614 &   9.563  \\ 
GAV& CLI & MIL &   17.216 &   3.894  & 	   1.338  & 112.469 & 1.803 &   5.462 &   1.227 &   3.669 &   69.440 &  13.204 &   3.302  \\ 
GAV& CLI & FER &   19.522 & 34.283  & 	 43.317  & 158.274 & 1.982 &   5.070 &   4.553 &   2.593 &   23.120 &  29.962 &  21.974  \\ 
GAV& CLI & KRO &  14.206 &   6.894  & 	   1.665  &   50.123 & 1.666 &   5.068 &   3.114 &   2.743 &   31.778 &    5.876 &    3.925  \\ 
GAV& CLI & CHA &  15.985 &   6.429  & 	 48.694  & 167.581 & 1.728 &   5.411 &   1.283 &   3.520 &   77.556 &  21.434 &  13.190  \\ 
GAV& CLI & MAS &  22.255 & 18.741  & 	154.480 & 364.015 & 2.014 &   5.719 &   0.690 &   4.135 &  119.605 & 58.290 &  42.565  \\ 
\hline
\end{tabular}
\end{center}
\end{table*}
\begin{table*}
\begin{center}
\caption{Confidence levels for the 8 best models. For 
each combination {\sl lim}+ {\sl mas} + IMF, defined in columns 1, 2, and 
3, the confidence levels obtained when fitting separately each data set of 
observations to our models, for columns  4 to 14. Column 15 is the 
combined likelihood, $P_{11}$, calculated using all observational sets. 
Column 16 is $P_{7}$, eliminating the disc properties (stars, gas, and 
SFR radial distributions).}
\label{bestmodels}
\begin{tabular}{cccrrrrrrrrrrrrr}
\hline
{\sl lim} & {\sl mas} & IMF & $P_{\Psi_t}$ & $P_{[Fe/H]}$ & $P_{[\alpha/Fe]}$& $P_{MDF}$ & $P_{HI}$ & $P_{H_{2}}$ & $P_{*}$ & $P_{\Psi_R}$ & $P_{C/H}$ &$P_{N/H}$ &$P_{O/H}$  &$P_{11}$ &$P_{7}$ \\
\hline
VHK & CLI & KRO & 0.997 & 0.913 & 1.000 & 0.502 & 1.000 & 1.000 & 0.994 & 1.000 & 0.997 & 0.999 & 1.000 & 0.930  &  0.894 \\
MAR & CLI & KRO & 0.997 & 0.972 & 0.995 & 0.770 & 1.000 & 1.000 & 0.994 & 1.000 & 0.799 & 0.443 & 1.000 & 0.885  &  0.826 \\
GAV & CLI & KRO & 0.996 & 0.881 & 0.994 & 0.267 & 1.000 & 1.000 & 0.992 & 1.000 & 0.488 & 1.000 & 1.000 & 0.819  &  0.732 \\
KAR & CLI & KRO & 0.998 & 0.993 & 1.000 & 0.108 & 1.000 & 1.000 & 0.994 & 1.000 & 0.998 & 1.000 & 1.000 & 0.816  &  0.727 \\
MAR & HEG & MIL & 0.729 & 0.968 & 0.129 & 1.000 & 1.000 & 1.000 & 1.000 & 1.000 & 0.383 & 1.000 & 1.000 & 0.737  &  0.619 \\
MAR & HEG & KRO & 0.998 & 0.729 & 0.410 & 0.379 & 1.000 & 1.000 & 0.996 & 1.000 & 0.372 & 0.523 & 1.000 & 0.707  &  0.580 \\
VHK & PCB & FER & 0.950 & 0.697 & 0.970 & 0.035 & 1.000 & 1.000 & 0.983 & 1.000 & 0.533 & 1.000 & 1.000 & 0.668  &  0.532 \\
MAR & KOB & KRO & 0.996 & 0.530 & 0.825 & 0.069 & 1.000 & 1.000 & 0.993 & 1.000 & 0.995 & 0.265 & 0.999 & 0.644 & 0.501 \\
\hline
\end{tabular}
\end{center}
\end{table*}
This is the reason why the radial gradient is basically the same for the 
144 models, since we use the same SFR and infall history for all of 
them; that said, the absolute abundances change significantly, since, 
even using the same basic model, the combination {\sl stellar 
yields+IMF} may change the absolute values of abundances in the disc by 
a factor of 100 for C, more than a factor of 10 for N, and a factor of 
30 for O. Thus, the 144 models results show a dispersion clearly greater 
than the data and the comparison with data allows us to select the 
appropriate combination of yields and IMF.

This is the most important result of this section: that it should be 
possible to select, on the basis of our CEM, which of these combinations 
may be valid in reproducing the empirical data and which of them should 
be rejected.  This is an important point to note as, in order to 
reproduce a given observation which appears to show a higher or lower 
metallicity than predicted by a model, a common fall-back option is to 
invoke some mechanism(s) of mixing, enrichment, or dilution of 
abundances, to reconcile the discrepancy.  As we show here, the 
alternate suggestion that the correct selection of IMF or/and stellar 
yields may be an easier way to achieve the desired abundance patterns 
should not be dismissed.

\section{The selection of the best models}
\subsection{The application of  a $\chi^{2}$ technique}

The objective of this section is to find the best combination of {\sl 
IMF + stellar yields} able to reproduce the MWG data amongst the 144 
models computed and described in the above section.  In order to do 
this, we use a classical $\chi^{2}$ technique comparing the model 
results and the corresponding observational data, such as those used in 
Fig.~\ref{disk},~\ref{solar} and ~\ref{abun}. In Table~\ref{tabla_chi}, 
we give our $\chi^{2}$ results; for each model calculated with a 
combination of {\sl lim} set +{\sl mas} set and IMF, we show the 
$\chi^{2}$ obtained from the comparison of our models with the data for 
all observational sets we use.

As said before and shown in Fig.~\ref{disk}, all models are equally good 
at fitting the radial distributions of both phases of gas, stars, and 
the SFR. We confirm this fact with the values of $\chi^{2}$ for these 
quantities\footnote{In all cases, the value at $R=0$\,kpc has not bern 
used in these $\chi^{2}$ calculations, since the differences between 
data and models are large in this region, and thus our $\chi^{2}$ values 
would be biased toward models with high densities in the inner disk, 
regardless of the quality of the agreement elsewhere in the disc.} in 
Table~\ref{tabla_chi}. Basically for all models they fall below the 
limits corresponding to 80\% of confidence level; that is, models 
fulfill widely these constraints. Therefore, we analyse the fit of our 
models for the other 7 empirical datasets.

We have assumed that each model is represented by a $\chi^{2}$ 
distribution, and calculated the corresponding likelihood, $P_{i}$, or 
confidence level, (complement of the significance level $\alpha$ 
associated\footnote{$\alpha$ is the statistical significance, 
corresponding to a given $\chi^{2}$, giving the probability of rejecting 
the null hypothesis, given that it is true, the null hypothesis being 
that both sets (observations and model results) would represent the same 
sample.} to each $\chi^{2}$). The number of free parameters, $NF=3$ in 
all cases, and the number of points for the fitting, $N_{obs,i}$, 
variable for each data set $i$, give the number of degrees of freedom 
$k_{i}=NF-N_{obs,i}$.  The likelihood is calculated as

\begin{equation}
P_{i}=1-\alpha(\chi^{2}_{k_{i}}< x)=1-\int_{0}^{x}{\chi^{2}_{kÐ{i}}du}=\\
1-\int_{0}^{x}{\frac{u^{k_{i}/2}e^{-u/2}}{2^{k_{i}/2}\Gamma(k_{i}/2)}du}.
\end{equation}

After computing these likelihood values, we see that the SFR and 
enrichment histories, much like the C, N and O abundances, may be easily 
reproduced with some combinations of IMF+ yields, showing low values for 
$\chi^{2}$, and high likelihood $P_{i}$ values. However, the relation 
[$\alpha$/Fe]-[Fe/H] and the MDF are more difficult to fit, and thus 
constrain the selection of models able to reproduce simultaneously all 
data sets.

\begin{table*}
\begin{center}
\caption{Confidence levels for eight other best models selected without 
using the MDF. }
\label{bestmodels6}
\begin{tabular}{cccrrrrrrrrrrrrr}
\hline
{\sl lim} & {\sl mas} & IMF & $P_{\Psi_t}$ & $P_{[Fe/H]}$ & $P_{[\alpha/Fe]}$& $P_{MDF}$ & $P_{HI}$ & $P_{H_{2}}$ & $P_{*}$ & $P_{\Psi_R}$ & $P_{C/H}$ &$P_{N/H}$ &$P_{O/H}$  &$P_{10}$ &$P_{6}$ \\
\hline
 KAR & KOB & MIL & 0.915 & 0.998  &0.998 & 0.000 & 1.000 & 1.000 & 1.000 & 1.000&  1.000 & 1.000 & 1.000 &   0.991 & 0.985 \\
 KAR & CLI & MIL & 0.873 & 0.939  &0.919 & 0.000 & 1.000 & 1.000 & 1.000 & 1.000&  0.997 & 1.000 & 0.999 &   0.972 & 0.953  \\
 MAR & KOB & MIL & 0.908 & 0.993  &1.000 & 0.000 & 1.000 & 1.000 & 1.000 & 1.000&  0.747 & 1.000 & 1.000 &   0.961 & 0.936  \\ 
 KAR & PCB & FER & 0.958 & 0.800 & 0.977 & 0.009 & 1.000 & 1.00 & 0.982 & 1.000 & 0.880 & 0.996 & 1.000 & 0.957 & 0.932 \\
 MAR & CLI & MIL & 0.856 & 0.997  &0.999 & 0.000 & 1.000 & 1.000 & 1.000 & 1.000&  0.465 & 1.000 & 1.000 &   0.912 & 0.857  \\
 KAR & KOB & KRO & 0.997 & 0.661 & 0.976 & 0.002 & 1.000 & 1.000 & 0.992 & 1.000 & 0.532 & 0.999 & 1.000 & 0.898 & 0.836 \\
 VHK & PCB & FER & 0.950 & 0.697 & 0.970 & 0.035 & 1.000 & 1.000& 0.983 & 1.000 & 0.533 & 1.000 & 1.000 & 0.897 & 0.836\\
 VHK & KOB & KRO & 0.994 & 0.310 & 0.993 & 0.016 & 1.000 & 1.000 & 0.993 & 1.000 & 1.000 & 0.999 & 1.000 & 0.887 & 0.821 \\
\hline
\end{tabular}
\end{center}
\end{table*}

In order to choose the best models, we have computed the combined
likelihood, $P_{S}$
\begin{equation}
P_{S}=\left(\prod_{i=1,i\not=2}^{S}{P_{i}}\right)^{1/S},
\end{equation}
obtained as the geometrical average of the individual $P_{i}$ previously 
calculated for each data set, and $S$ is the number of used data sets. 
In this expression, we may assume that a good model is the one that 
simultaneously reproduces all data sets, including the ones pertaining to 
the present state of the disc, $\Sigma_{HI}$, $\Sigma_{H_{2}}$, $\Sigma_{*}$, 
and $\Sigma_{SFR}$; in that case, the number of datasets used is 
$S=11$. Conversely, we could only use the 7 datasets 
shown in Fig.~\ref{solar} and \ref{abun}, that is, the observed SFR 
and enrichment histories, the relation [$\alpha$/Fe]-[Fe/H], the MDF, 
and the radial profiles of C/H, N/H, and O/H.  Therefore, by maximising 
the combined likelihood $P_{11}$ or $P_{7}$, we are able to select the 
best models of our grid.  We have computed both values $P_{11}$ and 
$P_{7}$, and then, we have ordered our models by using the combined 
likelihood $P_{7}$ and taken the first 8 (which represents $\sim 5$\% of 
the total number of the calculated models), which have values 
$P_{7}$$>$$\sim$50\% . The order of models using $P_{11}$ is exactly the same 
for these models, showing values $P_{11}\gtrsim 65\%$. We show these 
models in Table~\ref{bestmodels}. Only four from our models present 
values higher than $\sim$70\% for the fit in the seven selected data 
sets (or $>80$\% in the entire set of observations) and all of them use 
CLI+KRO combinations. All the other models have $P_{7}< 50\%$ .

\begin{figure}
\includegraphics[width=0.38\textwidth,angle=-90]{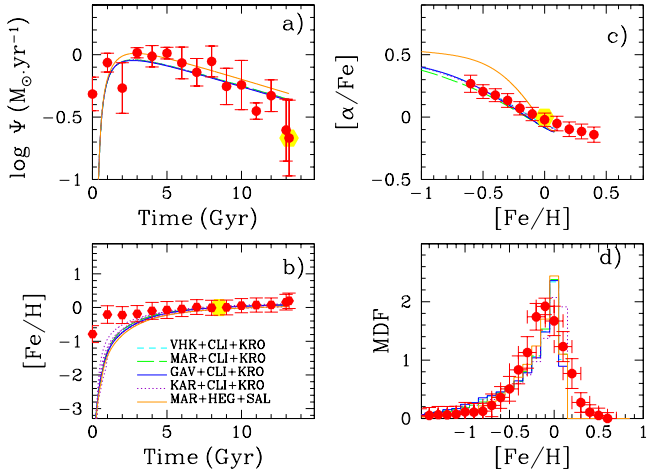}
\caption{Solar vicinity evolution of the best 4 models shown in
Table~\ref{bestmodels}: a) $\Psi(t)$, b) [Fe/H](t), c) [$\alpha$/Fe] {\sl
vs} [Fe/H] and d) MDF. Red and yellow dots have the same meaning as
in Fig.~\ref{solar}.  The orange dot-dashed line represents the model where
P$_{MDF}$=1 but P$\rm _{[\alpha/Fe]}=0$.}
\label{sv_bestmodel}
\end{figure}
\begin{figure}
\includegraphics[width=0.45\textwidth,angle=0]{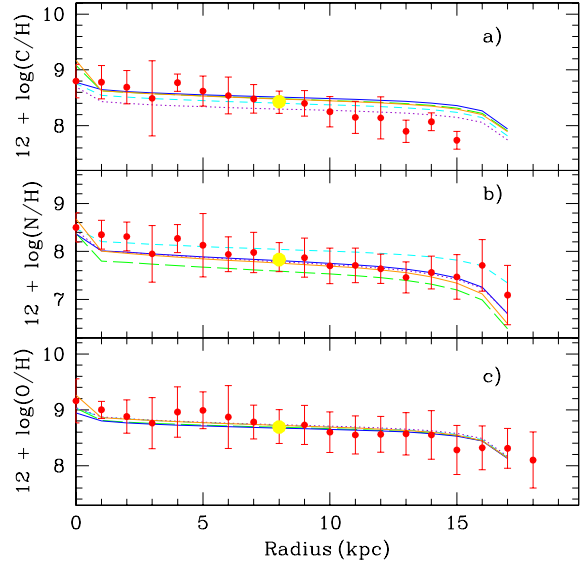}
\caption{Radial distributions of elemental abundances for the best 4
models compared with data. The meaning of colours, symbols, and types of
lines is the same as in Fig.~\ref{sv_bestmodel}.}
\label{abun_bestmodel}
\end{figure}

These results allow us to constrain the models, reducing the valid ones 
to only 4-8 models, depending on the {\sl goodness} we require. However, 
we must take into account that this conclusion is mainly due to the MDF, 
which show values of $\chi^{2}$ very high compared with most of the 
models. In fact, besides MAR+HEG+MIL, the 5$^{th}$ in the table, there 
there is only one o ther model, corresponding to the combination 
MAR+HEG+SAL, which has a value P$_{MDF}$=1.  However, these two models 
have a P$\rm _{[\alpha/Fe]}$=0.129 and 0, respectively, which implies 
they do not reproduce this relation at all.  Actually there is only one 
model, the first one of the table, with $P_{i} > 50\%$ for all columns. 
If we eliminate the MDF as a constraint for our models and calculate the 
equivalent $P_{6}$ and $P_{10}$, we find 11 models satisfying this 
condition for the ten other columns. Eight of them, shown in 
Table~\ref{bestmodels6}, using MIL or FER as IMF, are able to reproduce 
the six (or ten) remaining data sets within a confidence level $P_{6}$ 
higher than 80\% (or $\sim 90$\% for $P_{10}$).  In fact, models 1 to 4 
in Table~\ref{bestmodels}, showing $P_{11} > 80\%$, increase to values 
$P_{10}> 92\%$, when we don't take into account the columns 
corresponding to the MDF.  Therefore, models of this second table would 
also be valid, considering that many of the literature MDFs of the past 
decade should different maxima positions: \citet{casa11} found this 
maximum at $[Fe/H]\sim -0.05$, similarly to \citet{chan00, 
luck06,fur08}, while \citet{kor15} find it near $-$0.2\,dex, more in 
agreement with \citet{all04,nor04,hol07}.  Although we have used some of 
these datasets to obtain a bin-averaged MDF, it is likely that the error 
bars associated with these data are higher than the pure statistical 
ones included in our $\chi^{2}$ calculation.

Summarising, our best models are combinations of CLI-KRO with any {\sl 
lim} set. It is necessary to note that, given the possible uncertainties 
in the MDF, perhaps other combinations of stellar yields and IMF, as 
shown in Table~\ref{bestmodels6}, might succeed in reproducing the MWG 
data, mainly if other hypotheses pertaining to the evolutionary scenario 
(infall rate or SFR) are assumed.

\subsection{Results for the best models}

Having selected our best models, we plot their results in the subsequent 
figures to compare with the observational data.  In 
Fig.~\ref{sv_bestmodel}, we show the evolution with time of SFR, 
$\Psi(t)$, $\rm [Fe/H](t)$, [$\alpha$/Fe]-[Fe/H], and the MDF.  We have 
also drawn as an orange dot-dashed line the model MAR+HEG+SAL which does 
not reproduce the relation [$\alpha$/Fe]--[Fe/H] as said in the previous 
section.

Finally, we show in Fig.~\ref{abun_bestmodel} the elemental abundances 
of a) C, b), N, and c) O with the same line coding that in the previous 
Fig.~\ref{sv_bestmodel}. We see in the panel b) that model using VHK 
shows the highest N abundances of the four models, just within the limit 
of the uncertainties, while using MAR, with GAV lying between the two 
and closest to the date, as also found in \citet{gav06}. The radial 
gradient of O abundances obtained with recent data from \citet{henry10} 
and \citet{luck11} gives an averaged value of $-0.040$\,dex\,kpc$^{-1}$; 
C data gives a radial gradient of $\rm -0.048\,dex\,kpc^{-1}$, similar 
to the one for O.  For N we obtain a radial gradient of 
$-0.062$\,dex\,kpc$^{-1}$, slightly steeper than the one for C and O. 
The four models show radial distributions which seem in fair agreement 
with the observed radial gradients.

\section{Conclusions}

\begin{itemize}
\item By using our multiphase chemical evolution code, we have 
calculated 144 models applied to the MWG, with the same basic 
hypotheses, but different combinations of four low and intermediate mass 
stellar yield sets, with six massive stellar yield sets, and six IMFs.
\item We have analysed the observational data corresponding to the 
temporal evolution for SFR and iron abundance, the relative abundance 
[$\alpha$/Fe] as a function of [Fe/H], and the MDF for the solar region; 
further, we provided radial distributions of masses and elemental 
abundances at the present time for the Galactic disk, obtaining binned 
data sets averaged with different authors' samples.
\item Using a classical $\chi^2$ technique, we compared the results of 
our 144 models with the binned data points from the observational data.
\item Assuming that a good model is the one that simultaneously 
reproduces the observed SFR history, the [$\alpha$/Fe]-[Fe/H] 
relation, the MDF, and the radial profiles of C/H, N/H, and O/H, we 
defined a geometrical averaged likelihood from the product of the 
individual confidence levels for these 7 quantities.
\item We find that the best 4 of our 144 models are able to reproduce all 
observational data sets with confidence levels $P_{7}$ higher than 
$\sim$70\%, and use combinations CLI+KRO with any {\sl lim} yields. It is 
necessary to take into account that, given the possible uncertainties in 
the MDF, maybe other different combinations of stellar yields and IMF 
might be equivalently good to reproduce the MWG data, mainly if other 
assumptions regarding the infall rate or SFR are used.
\end{itemize}

\section{Acknowledgments}
This work has been supported by DGICYT grant AYA2010-21887-C04-02. Also, 
partial support from the Comunidad de Madrid under grant CAM 
S2009/ESP-1496 (AstroMadrid) is grateful.This work has been financially 
supported by the grant numbers 2010/18835-3, 2012/22236-3 and 
2012/01017-1, from the S\~{a}o Paulo Research Foundation (FAPESP).  This 
work has made use of the computing facilities of the Laboratory of 
Astroinformatics (IAG/USP, NAT/Unicsul), whose purchase was made 
possible by the Brazilian agency FAPESP (grant 2009/54006-4) and the 
INCT-A.  M.Moll\'{a} thanks the kind hospitality and wonderful welcome 
of the Jeremiah Horrocks Institute in Preston and of the Instituto de 
Astronomia, Geof\'{\i}sica e Ci\^{e}ncias Atmosf\'{e}ricas in Sao Paulo, 
where this work was partially done.  BKG acknowledges the support of the 
UK's Science \& Technology Facilities Council (ST/F002432/1, 
ST/H00260X/1, and ST/J001341/1). OC would like to thank H. Monteiro for 
enlightening discussions. We acknowledge the anonymous referee for very 
helpful comments.

\appendix

\section{Observational data}
\label{obs}

The observational data against which our CEMs are compared are now 
outlined.  These include the solar neighbourhood's temporal evolution, 
in addition to the present state of the disc, including the radial 
distributions of surface densities for stars, gas, and star formation 
rate, and elemental abundances of C, N, and O. Other data, such as 
[X/Fe], are usually represented as a function of [O/H] or [Fe/H], with 
the latter typically being employed as a proxy for time. Thus, we have 
also used the [$\alpha$/Fe] {\sl} -- [Fe/H] data of the solar vicinity 
to compare with our models.

\subsection{The Solar Vicinity}
\label{obs_sun}
\begin{table}
\caption{Binned SFR and  metallicity evolution for the solar vicinity}
\label{SV-binned}
\begin{tabular}{cccccc}
\hline
Time &  log\,$\Psi$ & $\Delta$($log \Psi$) & [Fe/H] & $\Delta$([Fe/H]) \\
(Gyr) &  \multicolumn{2}{c}{($\rm M_{\sun}\,yr^{-1}$)}&  & \\
\hline
    0   &  -0.3143 &   0.1667     &      -0.788   &      0.25 \\ 
    1   &  -0.0631 &   0.0935     &      -0.208   &      0.26 \\ 
    2   &  -0.2668 &   0.2510     &      -0.225   &      0.25 \\ 
    3   &   0.0165 &   0.0507     &      -0.184   &      0.24 \\ 
    4   &  -0.0118 &   0.1055     &      -0.097   &      0.25 \\ 
    5   &   0.0127 &   0.0577     &      -0.073   &      0.25 \\ 
    6   &  -0.0651 &   0.1568     &      -0.040   &      0.23 \\ 
    7   &  -0.1400 &   0.1372     &       0.017   &      0.22 \\ 
    8   &  -0.0530 &   0.1534     &      -0.007   &      0.22 \\ 
    9   &  -0.2535 &   0.1797     &       0.005   &      0.22 \\ 
   10   &  -0.2420 &   0.2441     &       0.056   &      0.21 \\ 
   11   &  -0.4513 &   0.0809     &       0.074   &      0.21 \\ 
   12   &  -0.3280 &   0.1581     &       0.077   &      0.21 \\ 
   13   &  -0.6039 &   0.3078     &       0.160   &      0.23 \\ 
   13.2 &  -0.6676 &   0.3759     &       0.200   &      0.23 \\     
\hline
\end{tabular}
\end{table}

\begin{figure}
\includegraphics[width=0.35\textwidth,angle=-90]{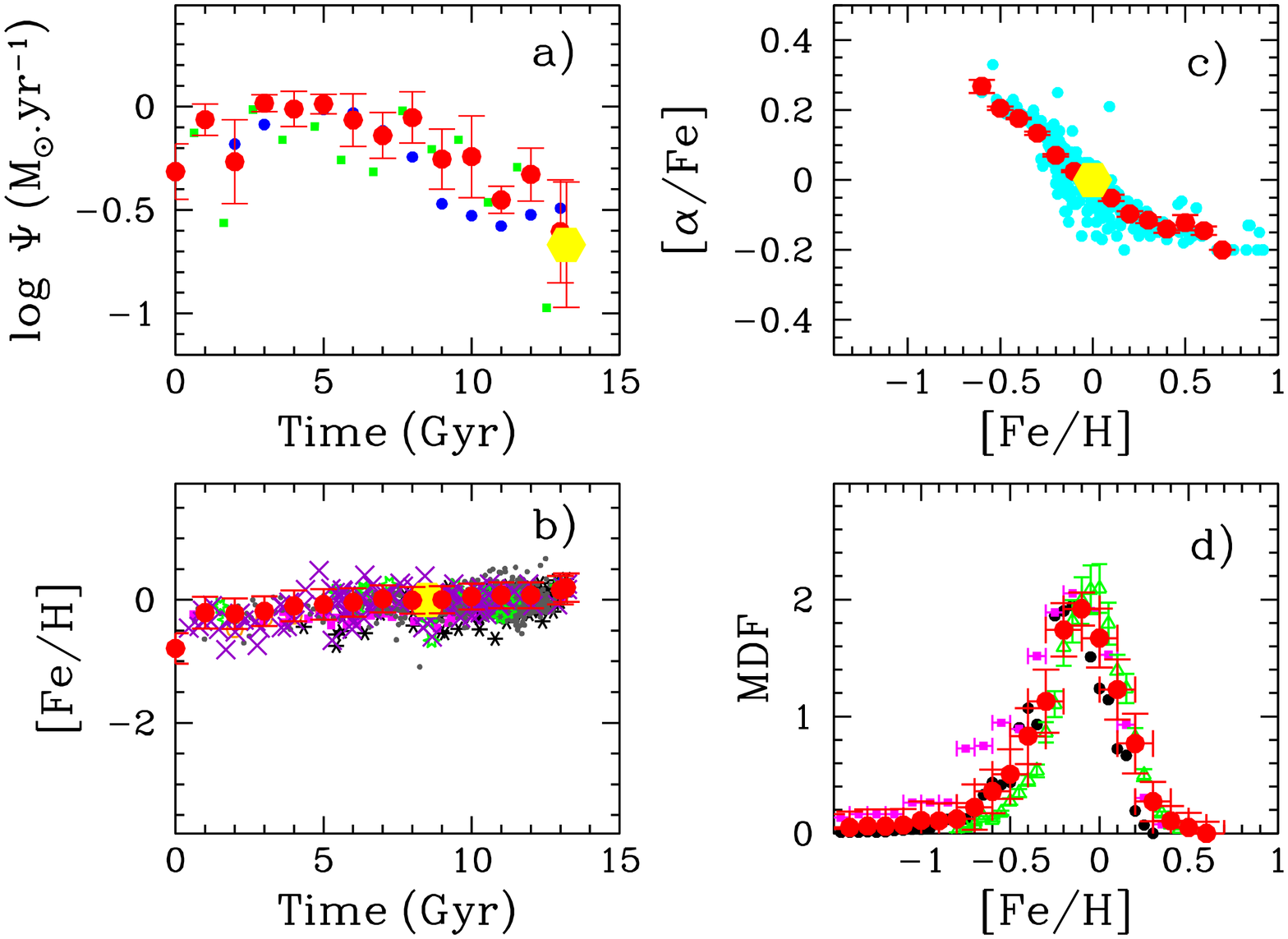}
\caption{Solar neighbourhood data: a) Star formation history $\Psi(t)$ 
with data from \citet{twa80} and \citet{rocha00b} as blue and green 
dots, respectively; and b) the age-metallicity relation 
$\rm{[Fe/H]}(t)$ with data from 
\citet{twa80,edv93,rocha00b,red03,casa11,ben14} and \citet{ber14} as 
orange open dots, black asterisks, blue full triangles, magenta full 
squares, grey small dots, green stars, and purple crosses, respectively. 
The large yellow dots are the solar neighbourhood
SFR at the present time in a) and 
the solar neighbourhood 
metallicity at the time when the Sun formed, 4.5\,Gyr 
ago, in b).  c) [$\alpha$/Fe] as a function of [Fe/H] from 
\citet{casa11} (cyan dots). The large yellow dot represents the solar 
abundances at coordinates
(0,0). d) The MDF with data from \citet{chan00,casa11} and 
\citet{kor15}(RAVE survey) as magenta squares, green triangles and black 
dots, respectively. In all panels the red dots with error bars are the 
binned results using all the noted datasets.}
\label{sun}
\end{figure}

\begin{table}
\caption{Binned [$\alpha$/Fe]-[Fe/H] relation and MDF for the solar vicinity}
\label{ofe-binned}
\begin{tabular}{ccccc}
\hline
[Fe/H] & [$\alpha$/Fe] & $\rm \Delta_{[\alpha/Fe]}$ &MDF & $\rm \Delta_{MDF}$ \\
\hline
-1.50 &       &       & 0.039  &  0.030 \\  
-1.40 &       &       & 0.062  &  0.040 \\  
-1.30 &       &       & 0.063  &  0.042 \\  
-1.20 &       &       & 0.013  &  0.050 \\  
-1.10 &       &       & 0.072  &  0.038 \\  
-1.00 &       &       & 0.129  &  0.050 \\  
-0.90 &       &       & 0.127  &  0.042 \\  
-0.80 &       &       & 0.190  &  0.100 \\  
-0.70 &       &       & 0.147  &  0.030 \\  
-0.60 & 0.268 & 0.019 &  0.411 &  0.088 \\  
-0.50 & 0.205 & 0.005 &  0.502 &  0.013 \\  
-0.40 & 0.176 & 0.002 &  0.886 &  0.123 \\  
-0.30 & 0.134 & 0.005 &  1.440 &  0.169 \\  
-0.20 & 0.071 & 0.004 &  1.900 &  0.065 \\  
-0.10 & 0.025 & 0.003 &  1.870 &  0.077 \\
 0.00 &-0.021 & 0.004 &  1.500 &  0.141 \\
 0.10 &-0.051 & 0.009 &  0.971 &  0.172 \\
 0.20 &-0.097 & 0.008 &  0.769 &  0.154 \\
 0.30 &-0.115 & 0.009 &  0.239 &  0.057 \\
 0.40 &-0.140 & 0.011 &  0.076 &  0.017 \\
 0.50 &-0.122 & 0.021 &        &        \\
 0.60 &-0.145 & 0.012 &        &        \\
 0.70 &-0.200 & 0.000 &        &        \\
 \hline
\end{tabular}
\end{table}

For the solar vicinities of our CEMs, we compare with extant 
observations pertaining to the time evolution of the SFR and the 
age-metallicity relation.  The SFR evolution is taken from \citet{twa80} 
and \citet{rocha00b}. In both cases, the data show a maximum around 8-10 
Gyr ago, that is, the onset of the SFR occurred at a time 3-5 Gyr after 
$t=0$, in agreement with more recent works of \citet{cig06} and 
\citet{rol13}.  These data are binned in 1~Gyr time-steps for the 
analysis which follows. The results, given in Table~\ref{SV-binned}, 
have then been normalised to the most recent values of the SFR in the 
solar region, corresponding to the final point at 13.2\,Gyr. The value 
for the present-day SFR for the entire MWG is estimated to be in the 
range [0.8-13] $\rm M_{\sun}\,yr^{-1}$ \citep{rana91}.  \citet{mis06} 
give a value of 2.7\,$\rm M_{\sun}\,yr^{-1}$, while \citet{cho11} find 
1.9\,$\rm M_{\sun}\,yr^{-1}$.  Taking into account the ratio of areas of 
the MWG disk within the optical radius and that of the solar region, we 
obtain a value of $\Psi_{\sun}\sim 0.266\,\rm M_{\sun}\,yr^{-1}$ 
for the region located at a galactocentric distance $R=8\,\rm kpc$, in excellent
agreement with the used value in \citet{cal10}.  
This value is the large yellow dot shown in Fig.~\ref{sun}a).  In this 
figure, we see the time evolution of the SFR once normalised to recover 
this value $\Psi_{\sun}\sim 0.266$ at 13.2\,Gyr. The values of $\Psi$ 
are given in Table~\ref{SV-binned}.

In panel b) of Fig.~\ref{sun}, we show the age-metallicity relation 
obtained with data from the literature as labelled, binned 
for each Gyr, as in panel a). Data from recent surveys such as RAVE 
\citep{boe13} or APOGEE \citep{and14} fall in the same region of the 
plane in panel b) when only the solar region\footnote{The solar region 
is defined as a 1~kpc annulus centered on a galactocentric radius of 
8~kpc, with a thickness of 200$-$500~pc.} data are selected. Our binned 
results are shown in Table~\ref{SV-binned}. They have been normalised to 
obtain a value [Fe/H]=$+$0 in $R=8\,\rm kpc$ at a time $t=8.5$\,Gyr, 
when the Sun was born, implying a shift of $+$0.1\,dex compared with the 
data of Fig.~\ref{sun}b). In both cases, we have added to the dispersion 
obtained from the binning process, a systematic error (representing 
observational uncertainties) of 0.05\,dex and 0.10\,dex in columns 3 and 
5, respectively.
\begin{table*}[h]
\begin{center}
\caption{Radial binned distributions obtained from observational data}
\label{binned}       
\begin{tabular}{ccccccccccccccc}
\hline
R & $\Sigma_{HI}$ &  error & $\Sigma_{H_{2}}$ & error & $\log{\Sigma_{*}}$ & error & $\log{\Sigma_{SFR}}$ & error & C/H & $\Delta C/H$& N/H & $\Delta N/H$ & O/H & $\Delta O/H$ \\
(kpc)   & \multicolumn{6}{c}{$\rm M_{\sun}\,pc^{-2}$} & 
\multicolumn{2}{c}{$\rm M_{\sun}\,pc^{-2}\,Gyr^{-1}$} &  & & & & \\
\hline 
  0    &    9.41   &  1.00  &  0.30 & 0.50 &          &         &     -0.370  &     0.15 & (8.66) & (0.30) &  8.39  &   0.31   &  9.02 & 0.40 \\
  1    &    3.97   &  1.88  &  3.82 & 4.90 &          &         &      0.603  &     0.52 & (8.64) & (0.30) & (8.24) &  (0.30)  & (8.86)& 0.30 \\
  2    &    2.37   &  2.04  &  5.18 & 5.39 &          &         &      0.706  &     0.47 & (8.55) & (0.30) & (8.20) &  (0.30)  & (8.74)& 0.30 \\
  3    &    2.39   &  2.12  &  3.48 & 2.24 &    2.43  &    0.01 &      0.983  &     0.59 &  8.35 &  0.67   &  7.84  &   0.59   & 8.62  & 0.45 \\
  4    &    3.86   &  2.35  &  5.69 & 3.35 &    2.50  &    0.13 &      1.163  &     0.35 &  8.63 &  0.15   &  8.16  &   0.29   & 8.82  & 0.45 \\
  5    &    5.06   &  2.14  &  8.28 & 2.87 &    2.40  &    0.07 &      1.185  &     0.24 &  8.48 &  0.27   &  8.02  &   0.66   & 8.83  & 0.33 \\
  6    &    5.04   &  2.06  &  8.47 & 1.67 &    2.25  &    0.07 &      1.181  &     0.27 &  8.40 &  0.32   &  7.83  &   0.36   & 8.77  & 0.56 \\
  7    &    5.44   &  1.58  &  4.59 & 1.72 &    2.09  &    0.08 &      0.963  &     0.26 &  8.34 &  0.25   &  7.87  &   0.41   & 8.69  & 0.30 \\
  8    &    5.69   &  2.38  &  3.15 & 1.42 &    1.95  &    0.09 &      0.723  &     0.29 &  8.28 &  0.20   &  7.77  &   0.30   & 8.56  & 0.30 \\
  9    &    7.69   &  2.13  &  2.44 & 0.80 &    1.79  &    0.10 &      0.594  &     0.25 &  8.26 &  0.22   &  7.76  &   0.41   & 8.60  & 0.35 \\
 10    &    6.52   &  2.18  &  1.96 & 1.18 &    1.69  &    0.14 &      0.510  &     0.43 &  8.11 &  0.27   &  7.59  &   0.37   & 8.45  & 0.36 \\
 11    &    6.16   &  2.06  &  1.24 & 0.80 &    1.51  &    0.03 &      0.403  &     0.39 &  8.01 &  0.29   &  7.60  &   0.36   & 8.41  & 0.34\\
 12    &    5.63   &  2.17  &  0.99 & 0.75 &    1.38  &    0.01 &      0.006  &     0.65 &  8.00 &  0.38   &  7.53  &   0.31   & 8.44  & 0.32 \\
 13    &    4.83   &  2.86  &  0.57 & 0.71 &    1.25  &    0.03 &      0.183  &     0.48 &  7.76 &  0.20   &  7.35  &   0.33   & 8.44  & 0.38 \\
 14    &    3.65   &  2.80  &  0.82 & 0.94 &    1.09  &    0.01 &     -0.260  &     0.57 &  7.93 &  0.17   &  7.45  &   0.34   & 8.42  & 0.43 \\
 15    &    2.96   &  2.69  &  1.09 & 1.84 &    0.94  &    0.01 &     -0.132  &     0.59 &  7.60 &  0.16   &  7.36  &   0.47   & 8.14  & 0.43 \\
 16    &    2.42   &  2.19  &  0.20 & 0.07 &    0.80  &    0.01 &     -0.520  &     0.15 &       &      &  7.60  &   0.54   & 8.14  & 0.39 \\
 17    &    2.15   &  2.17  &  0.13 & 0.05 &       &      &     -0.680  &     0.18 &    &      &  6.98  &   0.62   & 8.19  & 0.35 \\
 18    &    1.61   &  1.77  &  0.08 & 0.03 &       &      &     -0.890  &     0.15 &    &      &     &       & 7.96  & 0.50 \\
 19    &    1.18   &  1.66  &  0.03 & 0.01 &       &      &     -1.370  &     0.15 &    &      &     &       &    &    \\
 20    &    1.10   &  1.60  &    &    &      &      &          &       &    &      &     &       &    &    \\
\hline               
\end{tabular}  
\end{center}      
\end{table*}   
In panel c) we show the values of the $\alpha$-element abundances 
compared with those of iron, with data taken from \citet{casa11}. From 
the latter, we have selected those stars located between 7.5 and 
9.5\,kpc that lie within 0.5~kpc pf the mid-plane of the disc, for 
studying the evolution of the solar neighbourhood. These values are 
binned and shown in Table~\ref{ofe-binned}. In panel d) we show the 
metallicity distribution function (MDF) . Given the similarity of the 
three datasets, we have binned and normalised the result to unity,
and listed them in Table~\ref{ofe-binned}.
    
\subsection{The MWG disk: surface densities}
\label{obs_disk}

\begin{figure}
\includegraphics[width=0.35\textwidth,angle=-90]{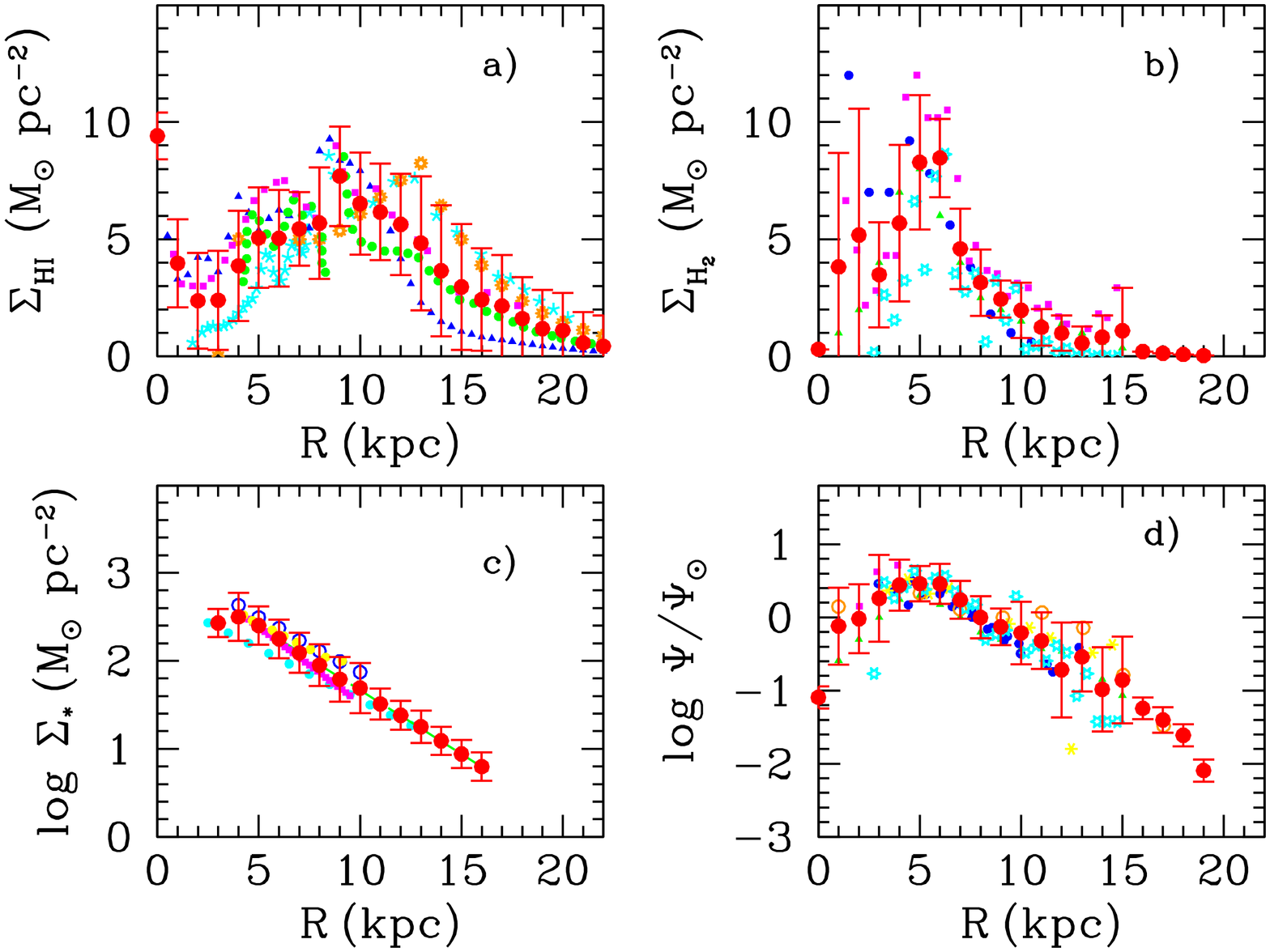}
\caption{Radial distributions of surface densities in the MWG for: a) 
diffuse gas, $\Sigma_{HI}$ (in $\rm M_{\sun}\,pc^{-2}$ units) with data 
from \citet{merr01,wol03,nak03,kal09} and \citet{pin13}, as cyan 
asterisks, orange stars, blue full triangles, green full dots, and 
magenta full squares, respectively; b) molecular gas, $\Sigma_{H_{2}}$ 
(in $\rm M_{\sun}\,pc^{-2}$ units) with data from 
\citet{wkee97,nak06,pin13} and \citet{urq14}, as green full triangles, 
blue full dots, magenta full squares, and cyan stars, respectively; c) 
Stellar profile $\Sigma_{*}$ (in $\rm M_{\sun}\,pc^{-2}$ units) with 
data from \citet{tal80,rana91,val00,bovy13,sof13}; d) the SFR surface 
density, $\Psi(R)/\Psi_{\sun}$, normalized to the Solar value 
$\Psi_{\sun}=0.266\,\rm M_{\sun}\,yr^{-1}$, estimated from 
\citet{mis06,cho11} in $R=8$\,kpc, with data from 
\citet{lac83,wkee97}, as green full triangles and blue full dots, and 
those taken from \citet{peek09} for pulsars, supernovae and H{\sc ii} 
regions, shown as yellow stars, orange open dots, and magenta full 
squares.  We have also used those from \citet{urq14}, represented by 
cyan asterisks.  These last two panels are in logarithmic scale.  The 
binned results are the large red dots with error bars in all of them.}
\label{mwg-dens}
\end{figure}

The radial gas distributions for both molecular and diffuse phases are 
well known. Since our model calculates separately both components, we 
also use these observations to fit our models. We use data from the literature,
shown in Fig.~\ref{mwg-dens} for diffuse HI and $\rm H_{2}$. By binning both 
sets, we obtain the results given in Table~\ref{binned} and shown in 
 panels a) and b). We see clearly a maximum around 10\,kpc for HI while 
H$_{2}$ shows an exponential shape from 4\,kpc to the outer disk. It 
also shows the well known molecular {\sl hole} inside $\sim$3\,kpc.

In panels c) of the same Fig.~\ref{mwg-dens} we also show the stellar 
surface density profile, including estimates from different authors as 
labelled. The most recent estimates for the solar stellar surface 
density give values between 33 and 64 $M_{\sun}$~pc$^{-2}$ 
\citep{kui89,kui91,val00,sie03,kho03,hol04,bien06,flyn06,web10,mil11, 
moni12,bur13,zhan13,bovy13}. These values depend on the scalelength for 
the disk $R_{d}$, which is in the range [2.15--4]\,kpc.  We show these 
data in panel c) of Fig.~\ref{mwg-dens} with our results after binning 
(red dots). In panel d), we show the SFR normalised to the solar value . 
The binned results for each kpc are also shown as red points, as in 
panels a), b), and c).

In Table~\ref{binned}, we present the resulting binned-averaged values 
of diffuse and molecular gas surface densities, and their associated 
errors, (columns 2 to 5), in $\rm M_{\sun}\,pc^{-2}$, for each radius 
given in column 1.  The stellar surface density profile is given, in 
logarithmic scale, with its associated error, in columns 6 and 7. In 
columns 8 and 9 we show the SFR surface density in $\rm 
M_{\sun}\,pc^{-2}\,Gyr^{-1}$.

\subsection{Disk elemental abundances for C, N, and O}
\label{abunobs}
\begin{figure}
\begin{center}
\includegraphics[width=0.43\textwidth,angle=0]{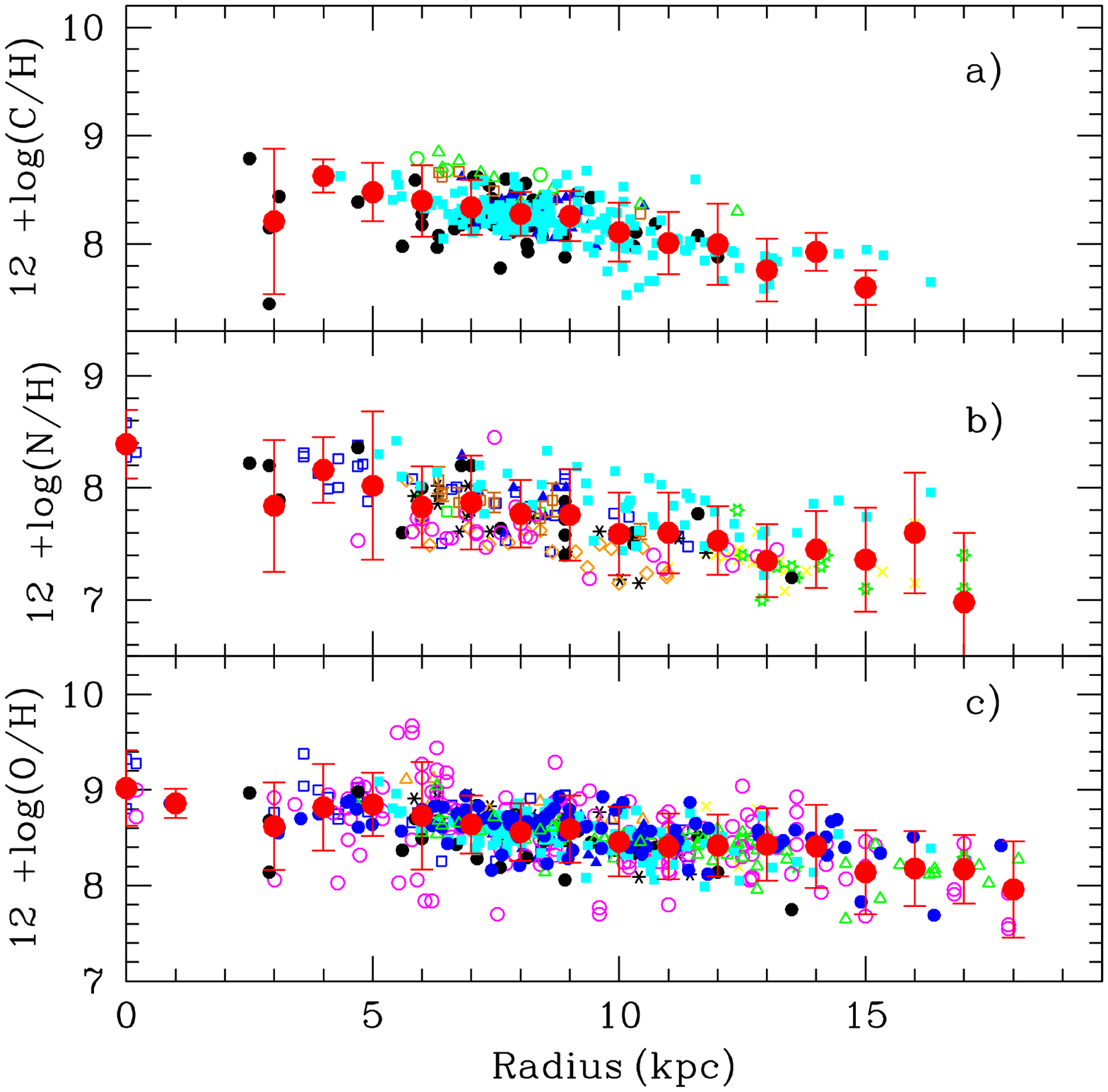}
\caption{Radial distributions of abundances (as $12+\log{(X/H)}$) for a)
C, b) N and c) O. Data are taken from the works noted in Table~\ref{authors},
where the symbol used for each is also given.  In all panels the
red full dots with error bars are the binned results obtained in this
work and given in Table~\ref{binned}.}
\label{mwg-abun}
\end{center}
\end{figure}

C, N and O abundances are the most important constraints for our models.  
Since N comes mostly from intermediate mass stars, O from the massive 
ones, and C from both, a fine-tuning of the stellar yields and IMF is 
necessary to reproduce simultaneously the three elements. We hope that 
any of the different combinations of yields from low and intermediate 
mass stars, with those from massive ones, with different IMFs, would 
give the right CNO elemental abundances.  We show in Fig.~\ref{mwg-abun} 
the three radial distributions for C, N, and O in panels a), b) and c), 
respectively. Data from different studies are plotted with different 
symbols, as listed in Table~\ref{authors}, while the red large dots are 
again our binned results (as $12+\log{(X/H)}$) shown in 
Table~\ref{binned}.

\begin{table}
\caption{List of data sources employed in Fig.~\ref{mwg-abun}.}
\begin{tabular}{lcccr}
\hline
Author & C  & N & O & Symbol\\
\hline
\cite{pei79} & -- & X & X & black $\ast$ \\
\cite{sha83} & --& X & X & orange $\diamond$\\
\cite{fs91}  & --& X & X & yellow $\times$\\
\cite{vies96} & --& X & X & green $\star$\\
\cite{aff97}  & --& X & X & blue $\blacksquare$\\
\cite{est99} &  X & X & X & green $\circ$\\
\cite{red03}&  X & X & X & blue $\blacktriangle$\\
\cite{daf04} & --& X & --& magenta $\circ$ \\
\cite{est05} & X & X & --& brown $\square$\\
\cite{gav06} & X & X & X & black $\bullet$\\
\cite{rud06} & --& --& X & magenta $\circ$ \\
\cite{henry10} & --& --& X & blue $\bullet$ \\
\cite{bal11} & -- & -- & X & green $\triangle$\\
\cite{luck11}& X & X & X & cyan $\blacksquare$\\
\cite{est13}& X & -- & --& green $\triangle$ \\
\hline
\end{tabular}
\label{authors}
\end{table}

\label{lastpage}
\end{document}